\documentclass[journal,twoside]{IEEEtran}

\usepackage[noadjust]{cite}
\usepackage{graphicx}
\usepackage{amsmath,amssymb,amsfonts}
\usepackage{algorithm}
\usepackage{algpseudocode}
\algrenewcommand\algorithmicdo{}
\usepackage{tabularx}
\usepackage{booktabs}
\usepackage{graphicx}
\usepackage{textcomp}
\usepackage{graphicx}
\usepackage{url}
\usepackage{multirow} 
\usepackage{rotating}
\usepackage{subcaption} 
\usepackage{flushend}
\usepackage{bm}

\DeclareMathOperator*{\argmin}{argmin}
\DeclareMathOperator*{\argmax}{argmax}

\newcommand{\x}{\mathbf{x}}
\newcommand{\y}{\mathbf{y}}

\renewcommand{\x}{\mathbf{x}}

\newcommand{\q}{\mathbf{q}}
\renewcommand{\u}{\mathbf{u}}

\newcommand{\w}{\mathbf{w}}
\newcommand{\m}{\mathbf{m}}
\newcommand{\Tone}{\text{T1}}
\newcommand{\Ttwo}{\text{T2}}
\newcommand{\PD}{\text{PD}}
\newcommand{\Dn}{\text{DAE}}
\newcommand{\Lcal}{\mathcal{L}}
\newcommand{\Bcal}{\mathcal{B}}

\newcommand{\Ncal}{\mathcal{N}}
\newcommand{\Acal}{\mathcal{A}}
\newcommand{\Mcal}{\mathcal{M}}
\newcommand{\Rcal}{\mathcal{R}}
\newcommand{\Ucal}{\mathcal{U}}
\newcommand{\Rbb}{\mathbb{R}}
\newcommand{\Cbb}{\mathbb{C}}

\newcommand{\Dq}{D^\mathcal{Q}}
\newcommand{\Dw}{D^\mathcal{W}}
\newcommand{\Bq}{\mathcal{B}^\mathcal{Q}}
\newcommand{\Bw}{\mathcal{B}^\mathcal{W}}
\newcommand{\BLCq}{\textsc{Bloch}^\mathcal{Q}}
\newcommand{\BLCw}{\textsc{Bloch}^\mathcal{W}}

\newcommand{\DM}{\textsc{d-match}}
\newcommand{\kcg}{K_{cg}}

\newcommand{\mriq}{MRI2Qmap}
\newcommand{\invivo}{\emph{in-vivo}}

\begin{document}

\title{MRI2Qmap: multi-parametric quantitative mapping with MRI-driven denoising priors}
\author{Mohammad Golbabaee, Matteo Cencini, Carolin Pirkl, Marion Menzel, Michela Tosetti, and Bjoern Menze 
\thanks{MG is with University of Bristol (m.golbabaee@bristol.ac.uk).  MC/MT are with IRCCS Fondazione Stella Maris. CP/MM are with GE Healthcare. MM is also with TH Ingolstadt. BM is with the University of Zurich.}
}

\maketitle
\begin{abstract}

Magnetic Resonance Fingerprinting (MRF) and other highly accelerated transient-state parameter mapping techniques enable simultaneous quantification of multiple tissue properties, but often suffer from aliasing artifacts due to compressed sampling. Incorporating spatial image priors can mitigate these artifacts, and deep learning has shown strong potential when large training datasets are available. However, extending this paradigm to MRF-type sequences remains challenging due to the scarcity of quantitative imaging data for training. \emph{Can this limitation be overcome by leveraging 
sources of training data from 
clinically-routine
weighted MRI images?}
To this end, we introduce \mriq, a 
plug-and-play quantitative reconstruction framework that integrates the physical acquisition model with priors learned from deep denoising autoencoders pretrained on large multimodal weighted-MRI datasets. \mriq\ demonstrates that spatial-domain structural priors learned from independently acquired datasets of routine weighted-MRI images can be effectively used for 
quantitative MRI reconstruction. 
The proposed method is validated on highly accelerated 3D whole-brain MRF data from both \invivo\ and simulated acquisitions, achieving competitive or superior performance relative to existing baselines without requiring ground-truth quantitative imaging data for training. By decoupling quantitative reconstruction from the need for ground-truth MRF training data, this framework points toward a scalable paradigm for quantitative MRI that can capitalize on the large and growing repositories of routine clinical MRI.

\end{abstract}


% ==============================================================
% = I N T R O D U C T I O N
% ==============================================================
\section{Introduction}
\label{sec:intro}
Quantitative MRI (qMRI) enables measurement of tissue biophysical parameters, such as T1 and T2 relaxation times, and provides objective imaging biomarkers for tissue characterization and disease monitoring~\cite{tofts2005quantitative}.  %seiberlich2020quantitative}. 
However, conventional qMRI is bottlenecked by long acquisition times, as multiple image series must be acquired and repeated for each parameter, making clinical adoptions challenging. 
% Despite it potential, conventional qMRI has seen limited clinical adoption due to prohibitively long scan-times, as multiple image series must be acquired and repeated for each parameter. 
Magnetic Resonance Fingerprinting (MRF)~\cite{ma2013mrf} and related transient-state methods~\cite{jiang2015fisp,gomez2020-3dqti,cao2022tgas} 
%add review papers about sequences?/cohen etc?
address this limitation by jointly encoding multiple tissue parameters using short dynamically-varying excitation sequences and highly undersampled k-space acquisitions.  While substantially reducing scan time, such acceleration can introduce pronounced aliasing artifacts, necessitating advanced algorithms to mitigate them for accurate quantitative mapping.

Methods for MRF reconstruction range from variational formulations that combine physical acquisition constraints with classical sparsity and low-rank image priors~\cite{asslander2018admm, mazor2018flor, golbabaee2021lrtv}, to more recent deep learning approaches that leverage data-driven priors and the strong representational capacity of neural networks, which now define the state of the art~\cite{fang2019supervisedmrf,mrf_tensor_unrolling23,delics2025deep,mrfdiph-miccai25}. 
Most existing methods rely on supervised, end-to-end training to map undersampled MRF data to high-quality, densely sampled quantitative maps. Their performance, however, critically depends on large training datasets with paired ground-truth parameter maps, which are difficult\textemdash if not impractical\textemdash to obtain due to prohibitively long scan times. Moreover, unlike conventional weighted MRI, qMRI acquisitions (accelerated or not) are not yet a clinical routine, resulting in limited dataset size and diversity. Reducing this reliance on quantitative training data therefore remains a central challenge.

Given the far broader availability of routine weighted-MRI data compared with quantitative MRI, and  the structural overlap 
between these imaging domains, we ask whether MRF reconstruction can benefit from priors learned from independently acquired weighted-MRI datasets. 
These imaging domains are related:  
multi-parametric quantitative maps, including those estimated by MRF, can be used to synthesize conventional weighted-MRI contrasts across multiple modalities via Bloch equations~\cite{warntjes2008rapidqmri, gulani2004synthmri, gomez2020-3dqti}. Building on this connection, we propose \mriq, a plug-and-play framework that integrates spatial priors learned from multimodal weighted-MRI datasets into 
% scan-specific 
MRF reconstruction 
from undersampled acquisitions. \mriq\ comprises three key components: (i) Bloch signal models to synthesize multiple weighted-MRI contrasts from estimated quantitative maps; (ii) restoration of these synthesized images using deep denoising autoencoders pretrained on large weighted-MRI datasets; and (iii) iterative updates of the quantitative maps that enforce joint consistency with the scan-specific physical acquisition model, measured data, and the spatially-restored synthesized images provided by the denoising priors. Our experiments on highly accelerated 3D whole-brain MRF acquisitions demonstrate that MRI-driven priors are effective in mitigating undersampling artifacts while capable of producing high-quality quantitative maps. 

\subsection{Related works}
\subsubsection{Model-based approaches}
Early MRF works~\cite{ma2013mrf,mcgivney2014svdmrf} employed Fourier back-projection to convert undersampled k-space data into image time-series, followed by dictionary matching, where voxels signal evolutions were compared to a dictionary of theoretical tissue fingerprints from Bloch equations, for quantitative mapping. Subsequent studies minimized a k-space loss while enforcing Bloch constraints through iterative dictionary matching~\cite{asslander2018admm,davies2014blip}. To reduce computational cost, approximate matching strategies~\cite{cauley2015fastgdm, golbabaee2019coverblip} and low-rank subspace models for dictionary compression~\cite{mcgivney2014svdmrf,asslander2018admm, zhao2018lowrank} were introduced. However, the absence of spatial regularization left these approaches prone to aliasing artifacts.
Later works e.g.~\cite{mazor2018flor, zhao2017simultaneous, lima2019sparsity,golbabaee2021lrtv,  cao2022tgas}  
included 
sparsity and low-rank regularization to reduce these artifacts; 
While effective, their performance usually lags behind deep learning methods, particularly at highly accelerated acquisitions. 

\subsubsection{Deep learning approaches}
\label{sec:lit-supdl} 
Early MRF deep learning works~\cite{hoppe2017dl4mrf,cohen2018drone,golbabaee2019geometry} used relatively small networks for voxel-wise parameter mapping after image reconstruction;  however, by processing voxels independently, these approaches lacked the spatial context necessary to mitigate correlated aliasing artifacts. 
In contrast, dominant recent literature incorporates spatial information by training on anatomical MRF maps using various architectures
including convolutional  networks~\cite{fang2019supervisedmrf, soyak2021convica, cheng2022_GCNN_3DMRF, delics2025deep}, 
 deep unrolling~\cite{golbabaee_miccai2020,mrf_tensor_unrolling23}, 
plug-and-play~\cite{fatania2022plug}, 
transformers~\cite{li2024_MRF_VIT},
and diffusion models~\cite{mrfdiph-miccai25}. Despite improved performance, these methods remain fundamentally constrained by the scarcity of 
MRF data and 
high-quality anatomical ground-truth, which limits their scalability.  
Some more recent studies also explored scan-specific (zero-shot) reconstructions, for the classical qMRI~\cite{bian2024_relaxmore}, multi-parametric imaging~\cite{zeroDeepSub}, and transient-state MRF~\cite{hamilton2022dipmrf,bardip_isbi24}, using only 
undersampled k-space data from one subject for training. While mitigating the need for ground-truth, these methods incur substantial computational cost, requiring \emph{de novo} network 
training 
for each scan, involving thousands of iterations for 2D reconstructions~\cite{hamilton2022dipmrf, bardip_isbi24,bian2024_relaxmore}, and several GPU-hours for 3D Cartesian acquisitions~\cite{zeroDeepSub} despite the lower computational burden of Cartesian imaging relative to non-Cartesian MRF. 
Moreover, prior work (e.g.,~\cite{chen2021equivariant}) suggests that effective reconstruction 
benefits from incorporating spatial-domain priors to compensate for information missing in  
highly undersampled measurements with large, non-trivial null spaces.

Our approach differs from the above paradigms by addressing quantitative data scarcity through priors learned from clinically ubiquitous, high-quality weighted MRI datasets, and by introducing an efficient algorithm that enables scan-specific, multicoil, 3D whole-brain non-Cartesian MRF reconstruction within minutes on a single GPU, without requiring qMRI/MRF training datasets or per-scan network retraining.

% ==============================================================
% = P R E L I M I N A R I E S
% ==============================================================
\section{Preliminaries}
\label{sec:preliminaries}
\subsection{The MRF inverse problem}
\label{subsec:prelim_mrf_problem}
MRF reconstruction is a nonlinear inverse problem: 
\begin{equation}
    \label{eq:forward_operator}
    \y \approx \Acal(\x)\; 
    \textrm{such that}\;\; \x_v=\Bq(\q_v), \;\forall v : \text{voxels}
\end{equation}
The goal is to estimate quantitative parameter maps (qmaps) $\q:=$ \{T1, T2, proton density (PD)\} over $n$ voxels. $\x \in \mathbb{C}^{s\times n}$ denotes the time series of magnetization images  (TSMI) to be reconstructed. 
The nonlinear MRF Bloch response model voxel-wise encodes qmaps into $s$-dimensional signal evolutions (fingerprints) of the TSMI, scaled by proton density:
\begin{align}
\label{eq:bloch_mrf}
\x_v = \Bq(\q_v):=\text{PD}_v. 
\BLCq(\text{T1}_v,\text{T2}_v)
\end{align}
The linear forward operator 
$\mathcal{A}$ 
relates TSMI to the undersampled k-space measurements $\y \in \mathbb{C}^{l\times cm}$ from $m$ spatial frequency locations across $c$ receiver coil channels, and $l$ time frames. It includes coil sensitivity profiles, a nonuniform FFT, and a linear dimensionality reduction ~\cite{mcgivney2014svdmrf,asslander2018admm, golbabaee2021lrtv} which is commonly used for computationally efficient reconstructions by compressing the time dimension of TSMI into adequately smaller $s\ll l$ time frames (coefficient channels). Reconstruction further relies on an MRF \emph{dictionary}, as a discretized approximation of \eqref{eq:bloch_mrf}, consisting of a lookup table ($\textsc{LUT}$) with $d$ sampled T1–T2 combinations and their precomputed Bloch responses $\Dq\in \mathbb{C}^{s\times d}$, where $ \Dq_j := \BLCq(j), \forall j\in \textsc{LUT}$. 
Projection of a TSMI $\tilde \x$ onto the Bloch constraint can be approximated by voxel-wise dictionary matching/search~\cite{davies2014blip}:  
\begin{align} 
\label{eq:standard_dm}
\q^{*}_v \leftarrow \DM(\tilde{\x}_v, \Dq) \approx \argmin\nolimits_{\q} \|\tilde{\x}_v-\Bq(\q)\|_2 
\end{align}
with $\DM(\tilde{\x}_v, \Dq) := \arg\min_{\PD,j\in \textsc{LUT}}\|\tilde{\x}_v -\PD.\Dq_j\|_2$.

%\vspace{-.2cm}
\subsection{Multimodal MRI synthesis}
A key advantage of multi-parametric maps produced by MRF-type acquisitions is their ability to synthesize conventional weighted-MRI images across multiple contrast modalities~\cite{warntjes2008rapidqmri, gulani2004synthmri, gomez2020-3dqti}.
Given quantitative maps $\q:=$\{T1, T2, PD\} common diagnostic contrasts (e.g., T1w, T2w, PDw) can be approximated via MRI Bloch signal models:
\begin{align}
\label{eq:bloch_mri}
\Bw(\q_v) := |\text{PD}_v| \cdot
\BLCw(\text{T1}_v, \text{T2}_v),
\end{align}
where $\Bw$ denotes a voxel-wise mapping from qmaps to MRI magnitude images $\m \in \Rbb^{s'\times n}$ across $s'$ modalities. For example, under a spin-echo signal model with synthetic acquisition parameters repetition (TR) and echo (TE) times:
\begin{align}
\label{eq:se}
|\PD_v|\exp\left(-\text{TE}/\Ttwo_v\right)\left(1-\exp(-\text{TR}/\Tone_v)\right),
\end{align}
T1w contrast arises from short TE, T2w from long TR, and PDw from short TE combined with long TR. Importantly, these contrasts retain useful 
dependencies on the underlying quantitative properties. 
In this work, 
we exploit this cross-domain relationship by synthesizing weighted-MRI contrasts from intermediate quantitative estimates, enabling MRI-driven image priors to iteratively mitigate undersampling artifacts and improve quantitative reconstruction in accelerated MRF. 

\vspace{-.2cm}
\subsection{Reconstruction with Plug-and-Play denoising priors}
\label{subsec:prelim_ddpm}
Given image samples from an unknown data distribution $p_x$, a denoising autoencoder $\Dn(.,\sigma^2)$  can be trained to remove additive Gaussian noise with variance $\sigma^2$ by minimizing
\begin{align}
\mathbb{E}_{x\sim p_x, \epsilon \sim \Ncal(0,\sigma^2 \mathbf{I})} \|\Dn(x+\epsilon, \sigma^2) -x\|_2^2.
\end{align}
The resulting model implicitly captures useful surrogate 
priors about the underlying data distribution. The denoising residual $\Dn(\tilde x, \sigma) -\tilde x \propto  \nabla_{\tilde x} \log p_{x,\sigma}(\tilde x)$ approximates
the score (gradient of log-likelihood) of the data distribution smoothed by a Gaussian convolution $p_{x,\sigma}:=p_{x}*\Ncal(\mathbf{0}, \sigma^2\mathbf{I})$~\cite{vincent-tweedie}. 
Thus, for sufficiently small $\sigma$, the DAE approximates a gradient-ascent step toward the manifold of plausible images. 
Denoising neural networks underpin modern plug-and-play (PnP) reconstruction methods~\cite{ahmad2020pnpmri, zhang2021drunet,kamilov2023plug}, 
which embed pretrained denoisers into iterative optimization algorithms 
by replacing regularization proximals (traditionally based on sparsity/low-rank) with expressive, data-driven DAE updates. PnP methods decouple the physical acquisition model from learned image priors, enabling scan-specific (zero-shot) reconstruction without retraining for different acquisition settings. Unlike prior PnP approaches trained on limited qMRI/MRF data~\cite{fatania2022plug,mrfdiph-miccai25}, we use DAEs pretrained on routine weighted-MRI datasets, and incorporate them into MRF reconstruction via MRI synthesis.

% ==============================================================
% = T H E   A L G O R I T H M
% ==============================================================

\vspace{-.1cm}
\section{MRI2Qmap}
\label{sec:the_algorithm}
We propose an augmented Lagrangian formulation for qMRI/MRF reconstruction integrated with MRI-driven priors:
\begin{align}
\label{eq:mri2q}
&\Lcal_{k}(\x,\q,\m,\u,\w) := f(\x) + \frac{\mu_k}{2} \left\|\x-\Bq(\q) + \u \right\|^2_2 \nonumber \\ 
&{} + \sum_{i\in \Mcal} \frac{\mu_k\gamma_i}{2} \left\|\m_i-\Bw(\q)_i + {\w_i}\right \|^2_2 + \alpha_i \Rcal_i(\m_i)
\end{align}
The goal is to reconstruct multi-parametric qmaps $\q$, the TSMI $\x$, and spatially-restored synthesized MRI images $\m=\{\m_i\}_{\i\in \Mcal}$ in contrast modalities $\Mcal=$\{T1w, T2w, PDw\}. $\u,\w$ are scaled Lagrange multipliers that are iteratively updated along with alternating minimization of $\Lcal_k$ over  $\x,\q,\m$ using ADMM algorithm iterations $k$~\cite{admm}. The first term $f(\x) :=\frac{1}{2}\|\y-\Acal \x\|_2^2$ enforces k-space consistent TSMI, the second term enforces TSMI-consistent qmaps via MRF Bloch response model~\eqref{eq:bloch_mrf}, and the third term enforces MRI-consistent qmaps via synthesis model~\eqref{eq:bloch_mri}. The regularizers 
$\Rcal_i$ impose modality-specific spatial priors on the synthesized MRI images, thereby restoring them and the linked qmaps from undersampling artifacts. Hyperparameters 
$\mu_k, \gamma_i, \alpha_i:= \lambda \gamma_i, \lambda > 0$ 
balance between terms and regularization.

 \vspace{-.2cm}
\subsection{The algorithm}
MRI2Qmap  (Algorithm 1) adopts a plug-and-play ADMM framework~\cite{pnp_wohlberg2013} with three key steps at each iteration: 
% \newline
\subsubsection{Updating k-space consistent TSMI}  minimizes~\eqref{eq:mri2q} over $\x$ while other variables fixed at their current estimates (Line 10): 
\begin{align} 
\x^k & \leftarrow prox_f\left(\Bq(\q)-\u, \mu_k^{-1} \right) \nonumber \\
& = \left(\Acal^H\Acal+\mu_k \mathbf{I}\right)^{-1} \left(\Acal^H \y +\mu_k (\Bq(\q) -\u ) \right).
\end{align} 
Here, $prox_h(\tilde \x,\tau) := \argmin_{\x}  \frac{1}{2} \|\x-\tilde \x\|^2 + \tau h(\x)$ denotes the 
proximal operator of a function $h$ with shrinkage parameter $\tau>0$.
For the least-squares data-fidelity term $f$, $prox_f$ reduces to solving a linear system, which can be updated  using a few conjugate-gradient (CG) iterations~\cite{ahmad2020pnpmri}. 

\subsubsection{MRI synthesis and  restoration} 
At each iteration, multimodal weighted-MRI images are synthesized from the current qmaps via $\Bw(\q)$ and restored spatially (Line 8) using a conditional MRI denoiser ($\Dn_\text{MRI}$) capable of operating across multiple specified MRI contrasts and Gaussian noise levels. The denoiser is trained on unpaired MRI datasets from diverse subjects, covering the contrast modalities 
$\Mcal$, and spatial resolutions and anatomies relevant to the MRF imaging task (see  Section~\ref{sec:implementation}). Following \cite{zhang2021drunet}, we employ an annealed denoising schedule with parameters $\sigma_k$,
%$\sigma_k:=\sqrt{\lambda/\mu_k}$,  
logarithmically spaced across iterations from $\sigma_1=\sigma_{\max}$ to $\sigma_K=\sigma_{\min}$, which  enables faster reconstructions 
% improves convergence and reconstruction efficiency 
(Section~\ref{sec:expe_ablation_anneal}). The denoiser serves as a plug-and-play replacement for the proximal operator of the regularizer in the $\m$-minimizing step of~\eqref{eq:mri2q} i.e., $prox_{\Rcal_i}\left( \Bw(\q)_i - \w_i, \sigma_{k}^2 \right)$ with $\sigma_k = \sqrt{\lambda/\mu_k}$, imposing %expressive, 
data-driven spatial priors on synthesized MRI modalities.

\subsubsection{Combined MRF-MRI dictionary matching} The $\q$-minimizing step updates the quantitative maps by enforcing joint consistency with the TSMI and the spatially-restored synthesized MRI images:
\begin{align}
\label{eq:q-update-bis}
\argmin\nolimits_\q \|\x+\u -\Bq(\q)\|^2_2 + \|\m + \w -\Bw(\q)\|^2_{\bm{\gamma}}.
\end{align}
Here, $\|x\|^2_{\bm{\gamma}}:=\text{trace}(x^H \text{diag}(\bm{\gamma})x)$  denontes a row-wise weighted squared norm for a matrix, where the weights 
% $ \bm{\gamma} := \{\gamma_i\}_{i\in \Mcal} \in \Rbb^{s'}$ 
$ \bm{\gamma} := \{\gamma_i\}_{i\in \Mcal}$ 
control the relative contributions of different MRI contrasts with respect to the TSMI in quantitative mapping. 
The objective is voxel-wise separable, 
which allows to 
extend dictionary matching 
to approximate this step.
Specifically, two dictionaries are precomputed from the Bloch models~\eqref{eq:bloch_mrf} and~\eqref{eq:bloch_mri}: the MRF dictionary $\Dq \in \Cbb^{s\times d}$ and the multimodal MRI dictionary $\Dw \in \Rbb^{s'\times d}$, both generated over a shared LUT of $d$ sampled T1-T2 pairs, where $\forall j\in \textsc{LUT}, \Dq_j := \BLCq(j)$ and $\Dw_j := \BLCw(j)$. 
Given a TSMI-MRI image pairs  $(\tilde \x , \tilde \m ) \in \Cbb^{s\times n} \times \Rbb^{s'\times n} $, the combined \emph{MRF-MRI dictionary matching}  (Line 9) is defined for each voxel $v$ as:
\begin{align*}
\q_v^* & = \DM\left((\tilde \x_v, \tilde \m_v), (\Dq,\Dw),\bm{\gamma} \right) \\
& :=\argmin_{j\in \textsc{LUT},\PD\in \Cbb} \|\tilde \x_v - \PD . \Dq_j \|^2_2 + \left\| \tilde \m_v - |\PD|. \Dw_j \right\|_{\bm{\gamma}}^2 
\end{align*}
with closed-form solutions (proof in Section~\ref{sec:app_dm_proof}):
\begin{align}
& (\Tone_v^*, \Ttwo_v^*)=j_v^*= \argmax_{j\in \textsc{LUT}} 
\frac{\left|\left\langle \Dq_j, \tilde \x_v \right\rangle \right| + \langle \Dw_j, \tilde \m_{v} \rangle_{\bm{\gamma}} } {\sqrt{ \| (\Dq_j\|^2_2 + \| \Dw_j\|^2_{\bm{\gamma}} }} \label{eq:fused-DM-search}, 
% \label{eq:optj}
\\
& \PD^*_v = \frac{\langle \Dq_{j^*}, \tilde \x_v \rangle}{|\langle \Dq_{j^*}, \tilde \x_v \rangle|} . \frac{\left|\left\langle \Dq_{j^*}, \tilde \x_v \right\rangle \right| + \langle \Dw_{j^*}, \tilde \m_{v} \rangle_{\bm{\gamma}} } {\| \Dq_{j^*}\|^2_2 + \| \Dw_{j^*}\|^2_{\bm{\gamma}}}. 
\label{eq:optpd}
\end{align}
Here, $\langle x, \tilde x \rangle_{\bm{\gamma}}:=\text{trace}(x^H \text{diag}(\bm{\gamma}) \tilde x)$ denotes a weighted inner product. 
Setting $\bm{\gamma}=\mathbf{0}$ recovers standard MRF-only dictionary matching as in~\eqref{eq:standard_dm} and~\cite{davies2014blip}, and disables the coupling between qmaps and MRI-domain priors. 
% discarding all MRI-domain priors.  
Following qmap estimation, updated TSMI and MRI images, $\Bcal^{\mathcal{Q}}(\q)$ and $\Bcal^{\mathcal{W}}(\q)$, are generated using their respective dictionary approximations.

Lines 11--12 update the ADMM Lagrange multipliers $\u, \w$, while Lines 7/13 obtain their scaled/unscaled forms according to the penalty parameters $\mu_k$, 
following~\cite[Chapter~3]{admm}. 
For initialization (Lines 1--5), we perform a few CG iterations to minimize the k-space loss $f(\x)$ to obtain $\x_\text{init}$, followed by standard MRF dictionary matching to obtain $\q_\text{init}$. 
Following \cite{zhang2021drunet}, Line 5 defines an increasing sequence for $\mu_k$ across iterations, determined by the user-specified annealed denoising schedule $\sigma_k$ and regularization parameter $\lambda$. 
Finally, because the MRF and MRI Bloch responses may differ 
%significantly 
in scale across modalities, we normalize the modality-wise weights $\bm{\gamma}$ by the relative norms of the TSMI and the synthesized MRI contrasts (Line 4) and use the normalized weights $\overline{\gamma}_{i\in\Mcal}\in[0,1]$ as algorithm input to control the MRF–MRI dictionary matching. % steps. 

\subsection{Implementation details}
\label{sec:implementation}

\subsubsection{Training the MRI denoiser}

Our approach was developed in PyTorch.\footnote{Source codes will be released upon acceptance.} The MRI denoiser $\Dn_\text{MRI}$  is a modality-conditional 3D residual UNet derived from the guided-diffusion toolbox~\cite{dhariwal2021ddpmbeatgans}. %\footnote{\url{https://github.com/openai/guided-diffusion}}. 
The model operates on 3D patches (size $64^3$) and conditions on MRI contrast (T1w, T2w, PDw) via class embeddings. The model is also conditioned on discretized noise levels 
%$\sigma \in \Sigma_\text{disc} \subset \Sigma$ 
through a noise-level embedding; specifically, 25 Gaussian noise levels $\Sigma$ logarithmically spaced within $[0.02,0.5]$ were used during training, spanning negligible noise to the upper effective range observed in our reconstruction experiments.
 The UNet employs five resolution levels with two residual blocks per level, channel widths $[64, 128, 256,  512, 512]$ from highest to lowest spatial resolutions, and attention layers (4 heads) at $16^3$ and $8^3$ feature resolutions. Training/inference was performed in fp16 mixed precision for improved computational efficiency.

The denoiser was trained on an unpaired MRI dataset comprising 3000 3D brain volumes assembled from 1000 T1w and 1000 T2w scans from the Human Connectome Project (1200 Subjects Release)~\cite{HCP2013}, and 500 PDw scans (sampled twice to balance modalities) from the IXI dataset~\cite{ixidataset}. 
All volumes were resampled to 1 mm$^3$ isotropic resolution, aligned to the MRF acquisition orientation, and intensity-normalized by scaling the 99.5th percentile to unity. 
Data augmentation 
% (via MONAI~\cite{monai2022}) 
(via MONAI~\cite{monai}) 
included 
smoothly varying bias-field distortions, 
random intensity scaling up to $\pm50\%$, and random cropping into $64^3$ patches at each iteration. 
Gaussian noise with standard deviation $\sigma\sim \Ucal(\Sigma)$ randomly drawn from discretized noise levels, was added to each training patch. Residual learning was adopted via the parameterization $\Dn_\text{MRI}(\tilde m,i,\sigma^2):= \tilde m -\epsilon_{\theta}(\tilde m, i, \sigma^2)$ where the network $\epsilon_\theta$ predicts the injected noise, given the noisy image $\tilde m$, modality label $i\in \Mcal$, and noise level $\sigma$. The network was trained to minimize an MSE loss between the true and predicted noise, using ADAM optimizer (learning rate $10^{-4}$, batch size 4) for 1000 epochs.

\subsubsection{Reconstruction parameters}
Experiments were conducted on a multicoil \invivo\ dataset and a single-coil simulated brain dataset using non-Cartesian MRF acquisitions from~\cite{cao2022tgas}. 

To accelerate the CG updates (Line 10), 
the $\Acal^H\Acal$ operations employed a Toeplitz approximation~\cite{baron2018toep} with spatiotemporal kernels~\cite{tamir2017t2, cao2022tgas}. 
Similar to~\cite{ahmad2020pnpmri} we found  that the CG subproblems do not require high-precision convergence (Section~\ref{sec:ablation_CG}). The number of ADMM iterations $K$ and inner CG iterations $\kcg$ depending on the conditioning of $\Acal^H\Acal$, were set to $K=\kcg=10$ in multicoil and $K=60, \kcg=20$ in single-coil experiments. 
For MRI synthesis $\Bw(\q)$, we used the spin-echo signal model in~\eqref{eq:se} with experimentally selected TR/TE settings of $(500 \text{ms}, 0), (\infty, 100 \text{ms})$ and $(\infty,0)$ for T1w, T2w, and PDw contrasts, respectively.
Additional TR/TE configurations were evaluated in Section~\ref{sec:ablation_TRTE}, with reconstruction performance remaining stable across relevant parameter ranges. 
The regularization parameter $\lambda$ was set to $2\times 10^{-6}$ for \invivo\ experiments and $10^{-6}$ for
simulated data. Equal modality weights ($\overline{\gamma}_i=0.3, \forall i\in \Mcal$) were used for dictionary matching. The base model employed exact dictionary search, with ablation results (Section~\ref{sec:ablation_gdm}) showing that approximate search can further optimize runtime. 

Annealed denoising schedules, logarithmically spaced between
$\sigma_1=0.2$ and $\sigma_K=0.02$ and rounded to the nearest levels in $\Sigma$, were used and provided best reconstruction runtime–accuracies (Section~\ref{sec:expe_ablation_anneal}). At each \mriq\ iteration, the three-modality MRI inputs to $\Dn_\text{MRI}$ were intensity-normalized, then shifted with random spatial offsets from $\Ucal([-32,32]^3)$ as a reconstruction-time augmentation. 
The shifted volumes were partitioned into non-overlapping $64^3$ patches, denoised, and reassembled, followed by inverse shifting and intensity rescaling to produce Line 8 outputs. 
Except denoising, other steps were performed in single precision.  

% %=========================
% % Algorithm FLow chart
% %=========================

\renewcommand{\algorithmicrequire}{\textbf{Input:}}
\renewcommand{\algorithmicensure}{\textbf{Output:}}

\begin{algorithm}[t!]
\caption{MRI2Qmap}
\label{alg:mri2qmap}
\begin{algorithmic}[1]

\Require 
  $\y$, $\Acal$, $\Dn_{\text{MRI}}$, $\Dq, \Dw$ (from $\Bq, \Bw$)
  \newline
  \textit{params:} $\{\sigma_{k}\}_{k=1}^K, \{\overline{\gamma_i}\}_{i\in \Mcal} , \lambda>0$, MRI modalities $\Mcal$ % e.g. $\Mcal=$\{T1w, T2w, PDw\}. 

% \Statex
% \textbf{Initialization}
% //Initialization
\State $\u=\w =\mathbf{0}$
\State $\x =\x_\text{init} =(\Acal^H\Acal)^{-1}\Acal^H \y$ 
\State $\q=\q_\text{init} =\DM(\x ,\Dq)$
\State $\bm{\gamma} := \left\{ \overline{\gamma}_i   \frac{\|\Bcal^{\mathcal{Q}}(\q)\|_2^2}{\|\Bcal^{\mathcal{W}}(\q)_i\|^2_2} \right\}_{i\in \Mcal}$ %$, \bm{\alpha}=\lambda \bm{\gamma}$
\State $\forall k:\; \mu_k:=\lambda/\sigma_k^2$ %, \;\; \bm{\zeta}_k := \mu_k\bm{\gamma}$

\For{$k = 1:K$}

\State $\u \gets \mu_k^{-1}\u, \;\; \w \gets \mu_k^{-1} \w$%\w \gets \text{diag}(\bm{\zeta}_k^{-1})\w $

\State $\forall i\in \Mcal:\; \m_i \gets \Dn_{\text{MRI}}(\Bw(\q)_i - \w_i, i, \sigma^2_{k})$ %$, \forall i\in\Mcal$

\State $\q \gets \DM\left((\x+\u, \m+\w), (\Dq,\Dw),\bm{\gamma} \right) $

\State $\x \gets \left(\Acal^H\Acal+\mu_k \mathbf{I}\right)^{-1} \left(\Acal^H \y +\mu_k (\Bcal^{\mathcal{Q}}(\q) -\u) \right)$

\State $\u \gets \u
    +\x -\Bcal^{\mathcal{Q}}(\q)$
\State $\w \gets \w
    +\m -\Bcal^{\mathcal{W}}(\q)$

\State $\u \gets \mu_k\u,\;\; \w \gets \mu_k \w$
    
\EndFor 
\newline
\Return
  $\q$ (qmaps), 
  $\x$ (TSMI), 
  $\m$ (synthetized MRIs).

\end{algorithmic}
\end{algorithm}

%=========================

%%%%%%%%%%%%%%%%%%%%%%%%%%%%%%%%%%
% TABLE with MAPES
%%%%%%%%%%%%%%%%%%%%%%%%%%%%%%%%%%
\begin{table*}[t!]
\scalebox{0.95}{
    \centering
    \setlength{\tabcolsep}{4pt}
    \setlength\extrarowheight{1pt}
    \begin{tabular}{llccc|c|ccc|ccc}
    \toprule 
         \multicolumn{2}{l}{\multirow{2}{*}{\textbf{Data/Method}} }
         & \multicolumn{3}{ c|}{\textbf{MAPE} (\%) $\downarrow$} & 
         {\textbf{CD}$\times 10^{-3}\downarrow$} &
         \multicolumn{3}{ c|}{\textbf{PSNR} (dB) $\uparrow$}& \multicolumn{3}{ c}{\textbf{SSIM} (\%) $\uparrow$} \\ \cline{3-12}
            &&\textbf{T1+T2} &\textbf{T1} & \textbf{T2}  
            &\textbf{TSMI} 
            &\textbf{T1w} & \textbf{T2w} & \textbf{PDw}  & \textbf{T1w} & \textbf{T2w} & \textbf{PDw} 
         \\    
\midrule
        \multirow{7}{*}{\rotatebox{90}{\emph{In-vivo}}}& SVDMRF &  44.40 & 14.97 (0.94) & 29.44 (2.22) 
        & -- & -- & -- & -- & --\\
         &ADMM & 23.14 & 9.40 (0.79)& 13.73 (1.27) 
         & 2.92 (0.50)& 32.18 (0.62) & 27.23 (0.65) & 27.81 (0.60) & 94.45 (0.78) & 91.35 (1.05) & 86.85 (1.20) \\
         & LLR &16.17 & 8.38 (0.59) & 7.79 (0.66)   & 3.25 (0.41) &
         32.93 (0.51)& 28.44 (0.59) &  29.09 (0.49) & 95.13 (0.68) &  92.97 (0.98)  &  86.43 (1.09)
         \\
         &LRTV   &14.18 & 6.53 (0.43) & 7.65 (0.58)
         & 2.71 (0.38) & 33.53 (0.46) & 28.84 (0.55) & 30.58 (0.45) & 97.02 (0.42) & 93.98 (0.72) & 90.65 (0.76)
         \\
         
         &\textbf{MRI2Qmap} &
        {12.01}
         & 4.93 (0.34)   & 7.08 (0.50)   & 0.85 (0.11) & 35.91 (0.45)& 30.03 (0.49)& 30.63 (0.52)& 97.95 (0.30)& 95.26 (0.58)& 91.42 (0.94) \\

         &ARNet$^*$ &
         12.82 & 5.61 (0.44) & {7.21} (0.63)   & 1.14 (0.20) & 35.83 (0.69) & 30.33 (0.76) & 31.74 (0.87) & 97.79 (0.33) & 95.31 (0.65) &  92.99 (0.92) \\

         &MRF-PnP$^*$ & {12.81} & {5.39} (0.51)&  7.42 (0.71)
         & 1.57 (0.33) & 36.33 (0.80) & 30.53 (0.81) & 32.37 (0.74) & 97.76 (0.43) & 95.37 (0.78) & 93.25 (0.81) 
         \\
         \midrule
         \midrule

         \multirow{7}{*}{\rotatebox{90}{\emph{BrainWeb}}}&SVDMRF & 254.95 & 70.79 (8.12) & 184.16 (25.66) 
        & -- & -- & -- & -- & -- & -- & --
        \\
         & ADMM & 36.08 & 12.74 (1.12) & 23.34 (2.24) 
         & 5.94 (1.02) & 26.47 (0.69)& 24.85 (0.61) & 26.29 (0.72)& 89.77 (0.75) & 88.60 (1.13) & 79.14 (0.99) \\

         &LLR &22.57 & 8.21 (0.48) & 14.36 (0.98)  
         & 3.31 (0.47)
         & 27.52 (0.46) & 27.20 (0.27) & 27.80 (0.62) & 90.39 (0.47) & 92.59 (0.50) & 81.69 (0.86) \\
         &LRTV & 19.22 & 7.32 (0.57)& 11.89 (1.08)  
         & 2.78 (0.46) & 29.16 (0.55) & 27.12 (0.58)& 29.16 (0.80) & 95.88 (0.50) & 93.93 (0.96) & 86.49 (1.15) \\

        &\textbf{MRI2Qmap} & 8.97 & 3.27 (0.15) & 5.70 (0.42) &0.61 (0.06)  & 32.33 (0.54) & 32.51 (0.27) & 32.98 (0.51) & 98.82 (0.10) & 97.96 (0.15) & 95.90 (0.53)
         \\
         %\midrule
         &ARNet$^*$  & 9.92 & 3.05 (0.14) & 6.87 (0.73) 
         & 1.02 (0.10) & 31.90 (1.06) & 27.37 (1.42) & 31.49 (1.34) &  97.99 (0.20) & 94.59 (2.13) & 94.53 (0.80) \\
         &MRF-PnP$^*$ &  8.25
         &  3.01 (0.09)
         & 5.24 (0.54) 
         & 0.56 (0.05) &  34.59 (0.88) & 33.84 (0.42) & 
         34.48 (0.76) &
         99.00 (0.08) & 98.46 (0.16) & 97.68 (0.23) \\
         \bottomrule
        
    \end{tabular}
    }
    \caption{{Mean (±SD) 
    % reconstruction
    performance on \invivo\ and BrainWeb datasets at $R=8$.
    $(^*)$ indicates methods trained with ground-truth MRF maps. 
    \mriq\ performs competitively with MRF-supervised baselines and outperforms remaining methods.} }
    \label{tab:baselines_combo}
    \vspace{-.1cm}
\end{table*}
%%%%%%%%%%%%%%%%%%%%%%%%%%%%%%%%%%%%
% MRI2Q vs ADMM
%%%%%%%%%%%%%%%%%%%%%%%%%%%%%%%%%%%%
\begin{figure*}[t]
    \centering
    \includegraphics[width=0.197\linewidth]{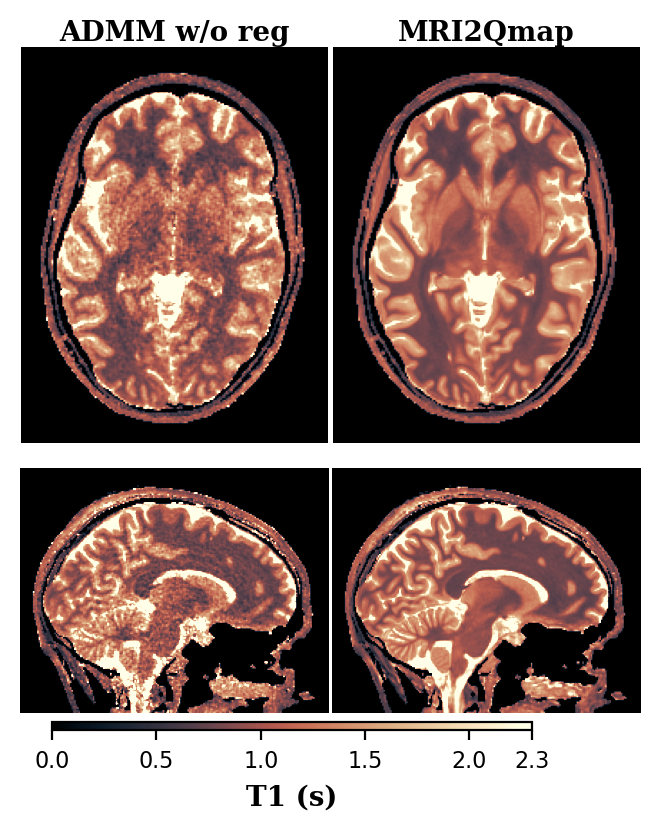}
    \includegraphics[width=0.197\linewidth]{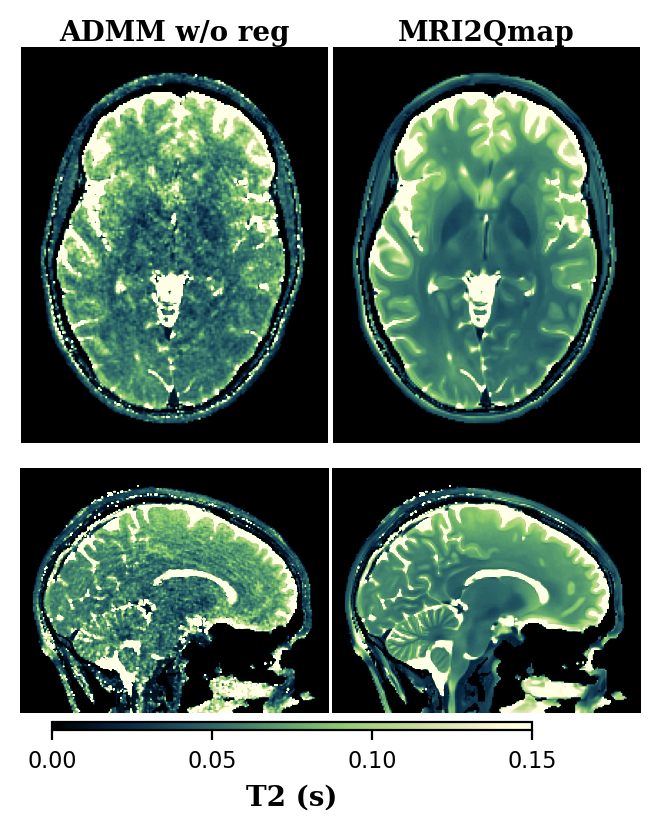}
    \includegraphics[width=0.197\linewidth]{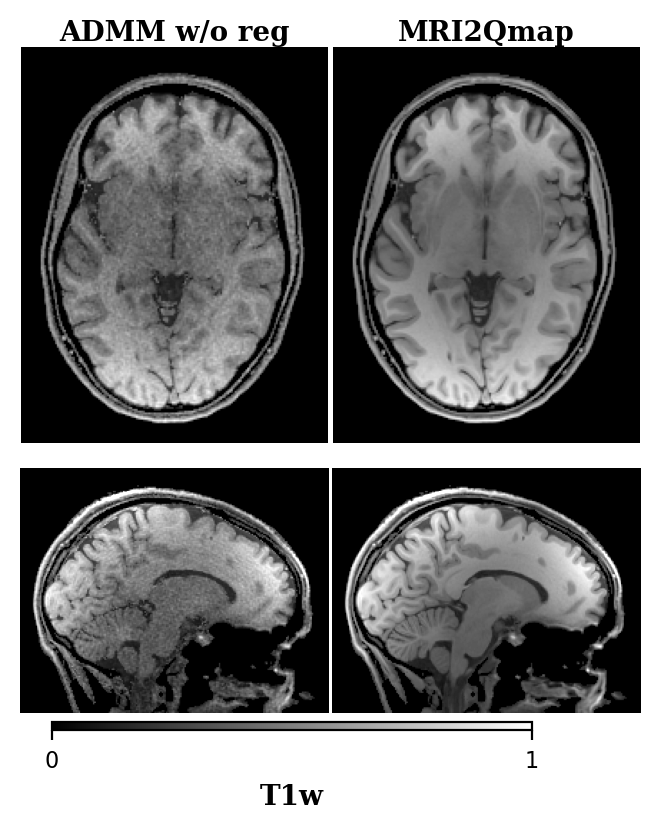}%
    \includegraphics[width=0.197\linewidth]{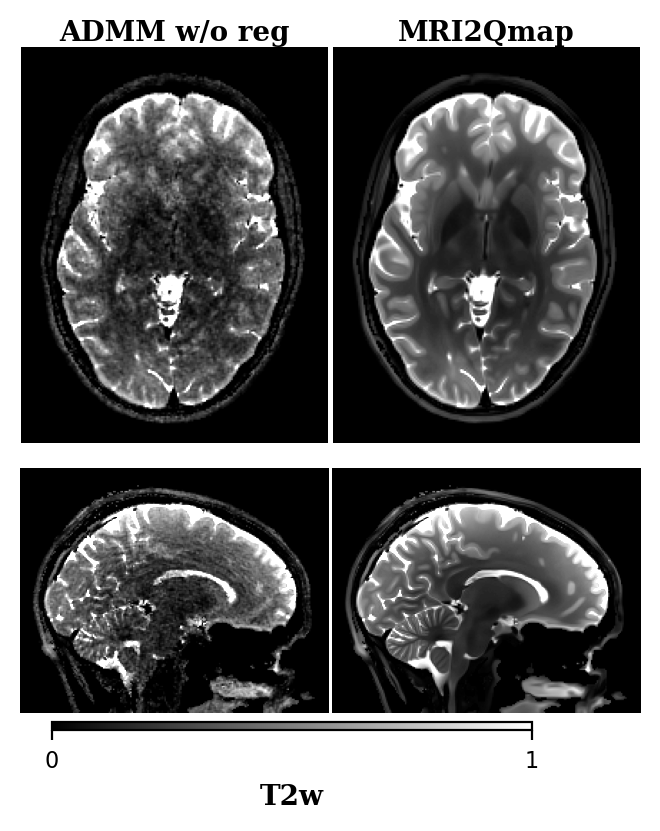}%
    \includegraphics[width=0.197\linewidth]{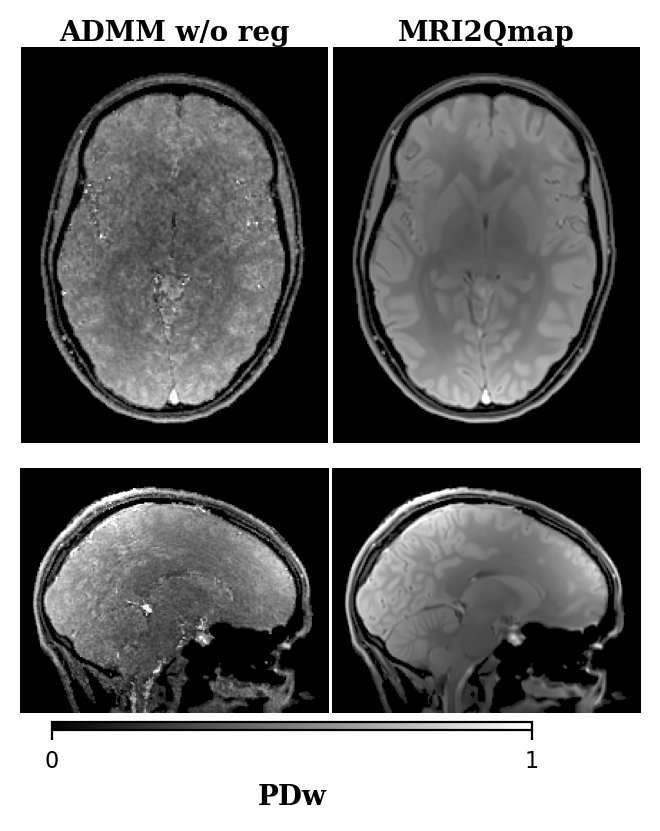}
    \caption{
    {MRI2Qmap vs. an ADMM reconstruction without spatial regularization. Quantitative T1/T2 maps and synthesized T1w, T2w, and PDw images are shown (electronic zoom recommended). By leveraging MRI-driven denoising priors, MRI2Qmap suppresses aliasing artifacts in synthesized contrasts and the underlying quantitative maps.} }
\label{fig:mri2qvsaddmm}
\vspace{-.3cm}
\end{figure*} 

%%%%%%%%%%%%%%%%%%%%%%%%%%%%%%%%%%%%
% MRI2Q vs baselines (VIVO)
%%%%%%%%%%%%%%%%%%%%%%%%%%%%%%%%%%%%
\begin{figure*}[t!]
    \centering
    \includegraphics[width=0.9\linewidth]{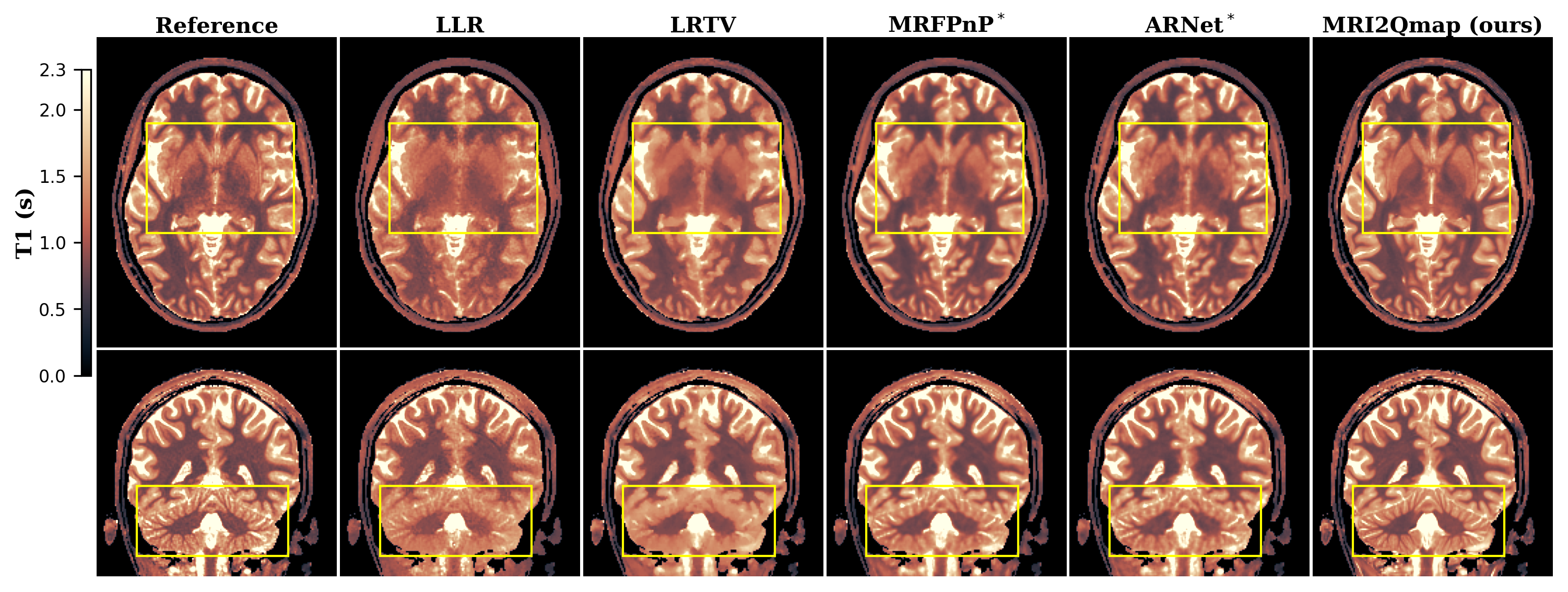}\\
    \vspace{-.2cm}\includegraphics[width=0.9\linewidth]{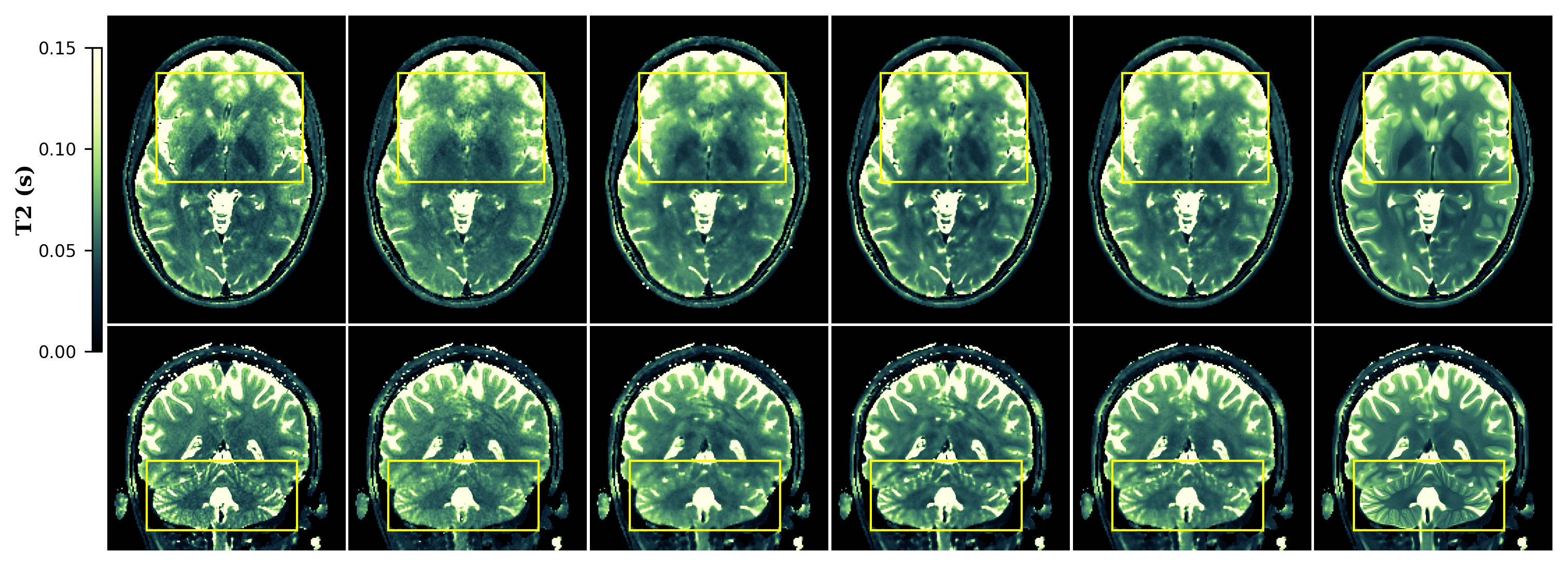}%
    \caption{
    {Reconstructed T1 and T2 maps at $R=8$ acceleration. 
    Relative to baselines ($^*$indicates MRF-trained), \mriq\ exhibits reduced artifacts and 
    improved anatomical delineation e.g.  
    for deep brain structures  
    (axial views) and the cerebellum (coronal views).  Electronic zoom in boxed regions is recommended. 
    }}
\label{fig:vivo-maps}
\vspace{-.3cm}
\end{figure*}
% ####====================================
% TSMI figure
% ####====================================

\begin{figure}[t!]
\centering 
\includegraphics[width=1.\linewidth]{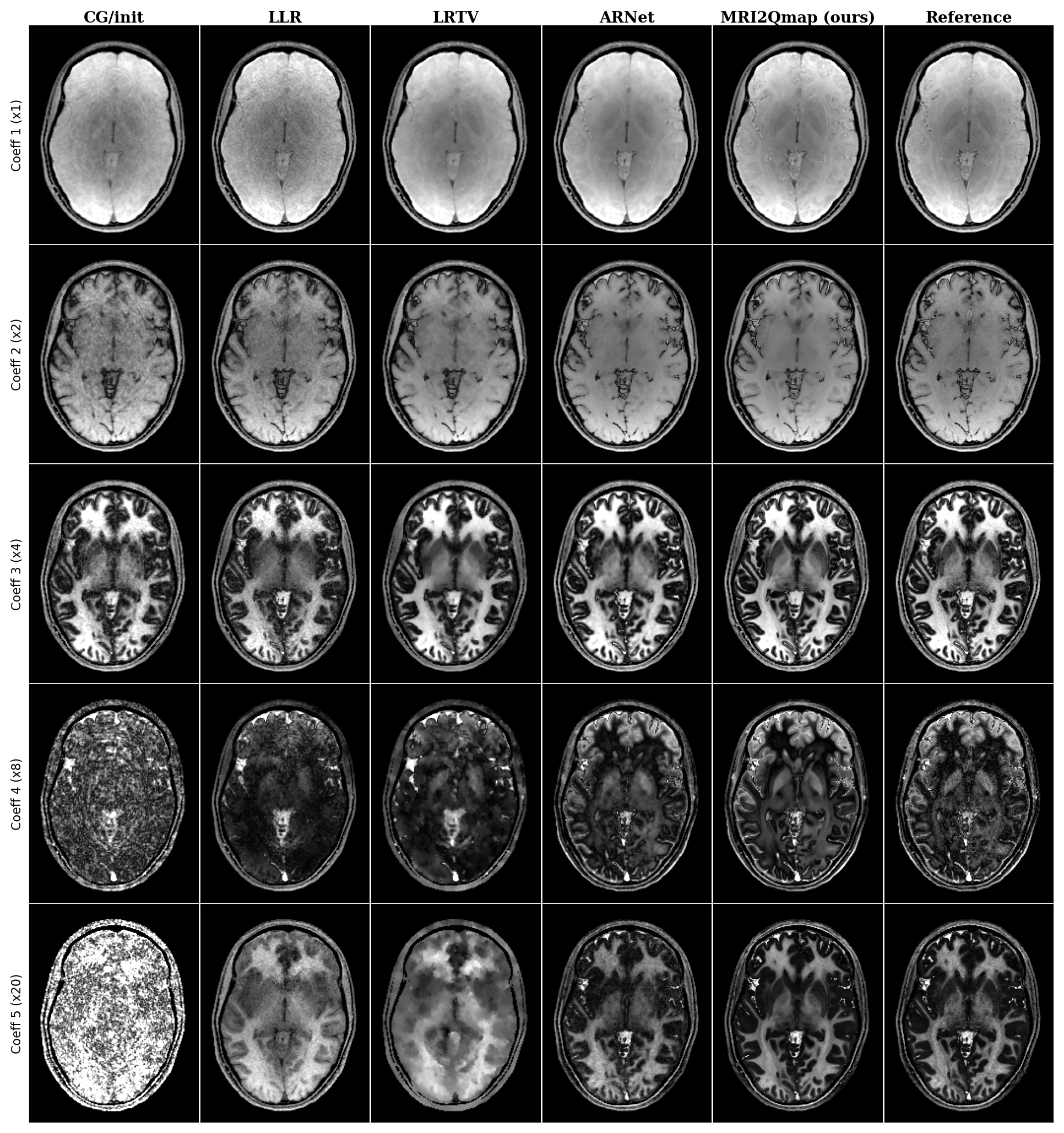}
    
\caption{TSMI magnitudes across $s=5$ compressed coefficient channels (rows) for the  subject in Fig.\ref{fig:vivo-maps},  
reconstructed by \mriq\ and baselines (columns). CG/init $\x_\text{init}$ initializes \mriq; 
ARNet was trained with MRF ground-truth data. 
}
\label{fig:tsmi-vivo-sub}
\vspace{-.3cm}
\end{figure}

\begin{figure}[t!]
    \centering
    \includegraphics[width=\linewidth]{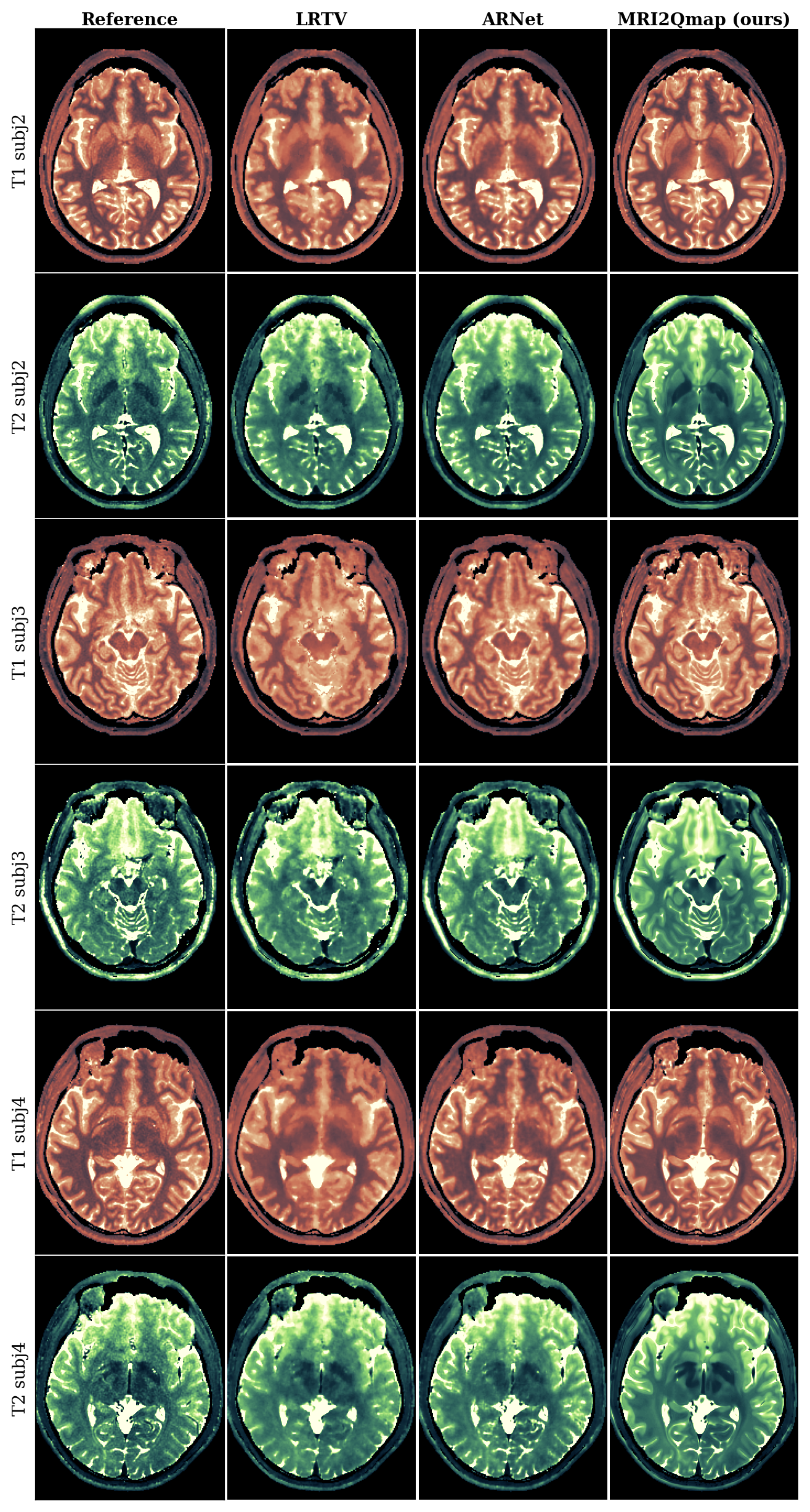}
    
    \caption{
    Reconstructed T1/T2 maps for additional \invivo\ subjects: \mriq\ vs. representative baselines (LRTV/ARNet). }
\label{fig:mri2q_vols}
\vspace{-.4cm}
\end{figure}

\section{Numerical Experiments}
\label{sec:experiments}
To evaluate the effectiveness of our method, we benchmark \mriq\ against representative MRF reconstruction baselines on two datasets: \invivo\ and simulated 3D brain acquisitions, both based on the TGAS–SPI–MRF protocol~\cite{cao2022tgas} with 8-fold k-space undersampling. All methods were implemented in PyTorch and run on a workstation with an Intel Core i9 CPU and a single NVIDIA RTX A6000 GPU. 

\subsection{Experimental setup}
\subsubsection{In-vivo data} 
We used 3D MRF brain data  from~\cite{delics2025deep}, acquired with the TGAS–SPI–MRF sequence on a 3T Premier scanner (GE Healthcare) using a 48-channel receive coil, at 1~mm$^3$ isotropic resolution and reconstructed to $256^3$ voxels. The gold-standard acquisition ($R=1$) used 48 readout groups (each sampling different k-space locations) to achieve dense k-space coverage within a 6-min scan. Each group consisted of an inversion pulse followed by a variable flip-angle schedule (10–75$^\circ$; TI/TE/TR = 20/0.7/12 ms) over $l=500$ timeframes with non-Cartesian 3D spiral k-space sampling trajectories. 
Following a similar scheme as~\cite{cao2022tgas}, 
six readout groups ($R=8$) were retrospectively subsampled to create highly accelerated sub-minute acquisitions. Data from 11 subjects were evaluated. Reconstructions from the densely sampled 6-min scans ($R=1$) served as reference, obtained via~\cite{asslander2018admm}  enforcing joint k-space and Bloch consistency. 
While our method does not require MRF training data, the tested deep learning baselines (ARNet, MRF-PnP) do. For these methods, we used a leave-one-subject-out strategy: 
for each test subject, the remaining 10 subjects were used for training, resulting in 11 separately trained models and a corresponding 11-fold evaluation. 

Pre-processing followed the toolbox in~\cite{delics2025deep}, including density compensation, field-of-view adjustment, compression to $c=10$ virtual coils, and sensitivity map estimation. The MRF dictionary ($\sim$26k atoms, $(T_1,T_2)\in[20,5000]\times[10,4000]$ ms) with SVD-based temporal compression from $l=500$ frames to $s=5$ coefficient channels~\cite{mcgivney2014svdmrf} was also adopted. 

\subsubsection{Simulated brain data}
Simulated data was generated from 3D qmaps (T1, T2, PD; 1 mm$^3$ resolution, $256^3$ voxels), 
% produced using MRtwin~\cite{MRtwin2025} and 
using 
MRtwin~\cite{MRtwin2025} and 
multi-subject anatomical brain models from BrainWeb database~\cite{brainweb} (see Appendix~S.II). 
% ~\ref{app:bw}). 
These qmaps served as ground truth and were used to create MRF time-series 
%(TSMI) 
data from the same TGAS–SPI–MRF sequence and Bloch dictionary. Gaussian noise at 25\% of the mean signal amplitude was added to each TSMI. 
Single-coil MRF measurements were simulated using the same 
$R=8$ k-space trajectories as in the \invivo\ study. 
%Using the same $R=8$ k-space trajectories as in the \invivo\ study, single-coil MRF measurements were simulated. 
Data from 10 subjects were used for evaluation; an additional 10 BrainWeb subjects were used to train deep-learning baselines (ARNet, MRF-PnP).

\subsubsection{Baselines}
\mriq\ with settings described in section~\ref{sec:implementation} was compared to the following baselines:  
\textbf{SVDMRF}~\cite{mcgivney2014svdmrf} uses gridding/adjoint operator $\x_\text{grid}=\Acal^H \y$
%to approximate inversion and 
to estimate the TSMI.  
%which is then passed to \DM\ for quantitative mapping. 
\textbf{MRF-ADMM}~\cite{asslander2018admm} uses ADMM for TSMI reconstruction by  enforcing k-space fidelity and Bloch/dictionary constraints i.e., the first two terms of \eqref{eq:mri2q}, without spatial regularization. 
%Spatial regularization was not used. 
For comparisons, reconstruction parameters ($\x_\text{init},K, \kcg,\mu$) were matched to those of \mriq. \textbf{LLR}~\cite{cao2022tgas} and \textbf{LRTV}~\cite{golbabaee2021lrtv} are convex baselines solving $\min_{\x} f(\x)+\Rcal(\x)$ with k-space fidelity and locally low-rank or total variation TSMI priors, respectively. Both methods use FISTA~\cite{fista}; LRTV applies the Chambolle–Pock proximal, %~\cite{CB2011}, 
while LLR uses singular-value soft-thresholding, with all proximal steps GPU accelerated. 
\textbf{Artifact Removal Network (ARNet)} 
is a supervised deep learning baseline trained to remove undersampling artifacts from initial TSMI reconstructions from $R=8$ acquisitions. Training used paired MRF data: inputs were randomly cropped $64^3$ patches from CG-reconstructed 
$R=8$ TSMI ($\x_\text{init}$ in Algorithm~\ref{alg:mri2qmap}, Line 2), and targets were the corresponding reference TSMIs derived from ground-truth qmaps (BrainWeb) or from long, densely sampled $R=1$ MRF acquisitions (\invivo).  
Several architectures have been proposed for this or closely-related tasks (see section~\ref{sec:lit-supdl});
For comparability, ARNet used the same UNet backbone as \mriq, with modified input/output channels. 
\textbf{PnP-MRF}~\cite{fatania2022plug}
is a plug-and-play ADMM method for TSMI reconstruction that combines k-space data fidelity with a TSMI denoising prior. Unlike \mriq, the denoiser operates directly on MRF timeframe data and was trained to remove simulated Gaussian noise from $64^3$ patches of reference TSMIs obtained via ground-truth qmaps (BrainWeb) or $R=1$ acquisitions (\invivo). Reconstructions from $R=8$ data alternated CG-based 
kspace fidelity 
with TSMI denoising, using same initialization and iterations count as \mriq.

For each ARNet/PnP-MRF baseline, twelve models were trained and evaluated: one for BrainWeb data, and eleven for \invivo\ subjects using leave-one-subject-out validation. Quantitative maps were obtained via standard MRF dictionary matching post hoc to all reconstructed TSMIs. 
All baselines used GPU acceleration and Toeplitz forward–adjoint approximations; 
see Supplementary Section~S.III for details.

\subsubsection{Evaluation Metrics}
T1 and T2 mapping performances were assessed using mean absolute percentage error (MAPE) over diagnostically relevant  white- and gray-matter regions: $\mathrm{MAPE}:=\mathbb{E}_{v\in M}\frac{|\hat{\q}_v-\q_v|}{|\q_v|}$,
where $\hat{\q},\q$ denote estimated and reference maps, $v$ indexes voxels, and $M$ is the anatomical mask. 
For BrainWeb experiments, masks were obtained directly from~\cite{brainweb}, whereas for \invivo\ data they were derived by segmenting reference images using~\cite{billot2023synthseg}.   
TSMI reconstruction quality was measured using cosine distance,
$\mathrm{CD}:=1-\mathbb{E}_{v\in M}\frac{|\langle\hat{\mathbf{x}}_v,\mathbf{x}_v\rangle|}{\|\hat{\mathbf{x}}_v\|_2\|\mathbf{x}_v\|_2}$.
Synthesized MRI images from predicted and reference qmaps were compared using PSNR and SSIM, averaged over 50 central axial slices, with air/background suppressed by thresholding low proton-density regions.

\subsection{Results}

Tables~\ref{tab:baselines_combo} compares \mriq\ with baseline methods on \invivo\ and BrainWeb datasets. SVDMRF shows the largest errors due to gridding-based inversion approximation (for this method, only T1/T2 mapping metrics are reported). MRF-ADMM improves upon this via iterative optimization; however, under highly undersampled acquisitions and without spatial regularization, its performance remains limited. LLR and LRTV further enhance reconstruction quality by incorporating spatial-domain priors; however, both are consistently outperformed by \mriq, which exploits learned MRI-domain priors. On the other hand, \mriq\ achieves 
results that are competitive with 
supervised MRF-domain baselines (ARNet and PnP-MRF), despite not being trained on ground-truth MRF data. Differences between \mriq\ and these baselines are generally small, with \mriq\ showing similar or in certain metrics marginally improved performance, e.g., for T1/T2	
and TSMI estimation for \invivo\ data.

Qualitative comparisons are further shown in Figs.~\ref{fig:mri2qvsaddmm}, \ref{fig:vivo-maps}, \ref{fig:tsmi-vivo-sub} for an \invivo\ test subject. 
Fig.\ref{fig:mri2qvsaddmm} compares \mriq\ and MRF-ADMM across reconstructed T1/T2 maps and the corresponding synthesized weighted MRI volumes;  
The two methods differ only in the inclusion of MRI priors in \mriq. 
While MRF-ADMM shows pronounced aliasing artifacts, \mriq\ effectively mitigates them, producing cleaned synthesized weighted images and, thanks to exploiting cross-domain MRF–MRI relationships, restored underlying T1/T2 maps.
Figure~\ref{fig:vivo-maps} extends the comparison to additional baselines (SVDMRF excluded due to poor performance). \mriq\ better preserves anatomical details and tissue boundaries, as seen in highlighted regions, including deep brain structures (e.g. putamen, pallidum, caudate) in axial views and the cerebellum in coronal views, for both T1/T2 maps. LLR/LRTV show noticeable over-smoothing, with residual aliasing particularly in T2 maps. Supervised MRF-domain baselines (e.g. ARNet) improve upon LLR/LRTV but tend to produce noisier maps with less well-defined tissue boundaries compared to \mriq, which may partly reflect the limited quality of their training targets from 6-min MRF acquisitions that are still somewhat undersampled and noisier than standard MRI data used for training \mriq.  
A similar trend is observed in the TSMI reconstructions (Fig.\ref{fig:tsmi-vivo-sub}), where the last three channels, the weakest and most challenging to recover, are better reconstructed by \mriq\ than the baselines, with clear improvement over the initialization $\x_{\text{init}}$, highlighting the effectiveness of %multimodal 
MRI-domain priors. 
Overall, these results demonstrate that \mriq\ achieves high-quality quantitative reconstructions without relying on ground-truth MRF training data. 
Additional reconstructions from other test subjects are shown in Fig.~\ref{fig:mri2q_vols}, with further \invivo\ and BrainWeb examples provided in Figs~S.1, S.2 and S.3.

\vspace{-.2cm}
\subsection{Discussion}
Here we study the impact of key \mriq\ design choices via ablation experiments, focusing on MRI denoising modalities, synthesis parameters, and runtime-related factors, followed by a discussion of limitations and future work.

\subsubsection{MRI denoising modalities}
\label{sec:ablation_switchoffmodal} 
\begin{figure}[t!]
\centering
\includegraphics[width=1.\linewidth]{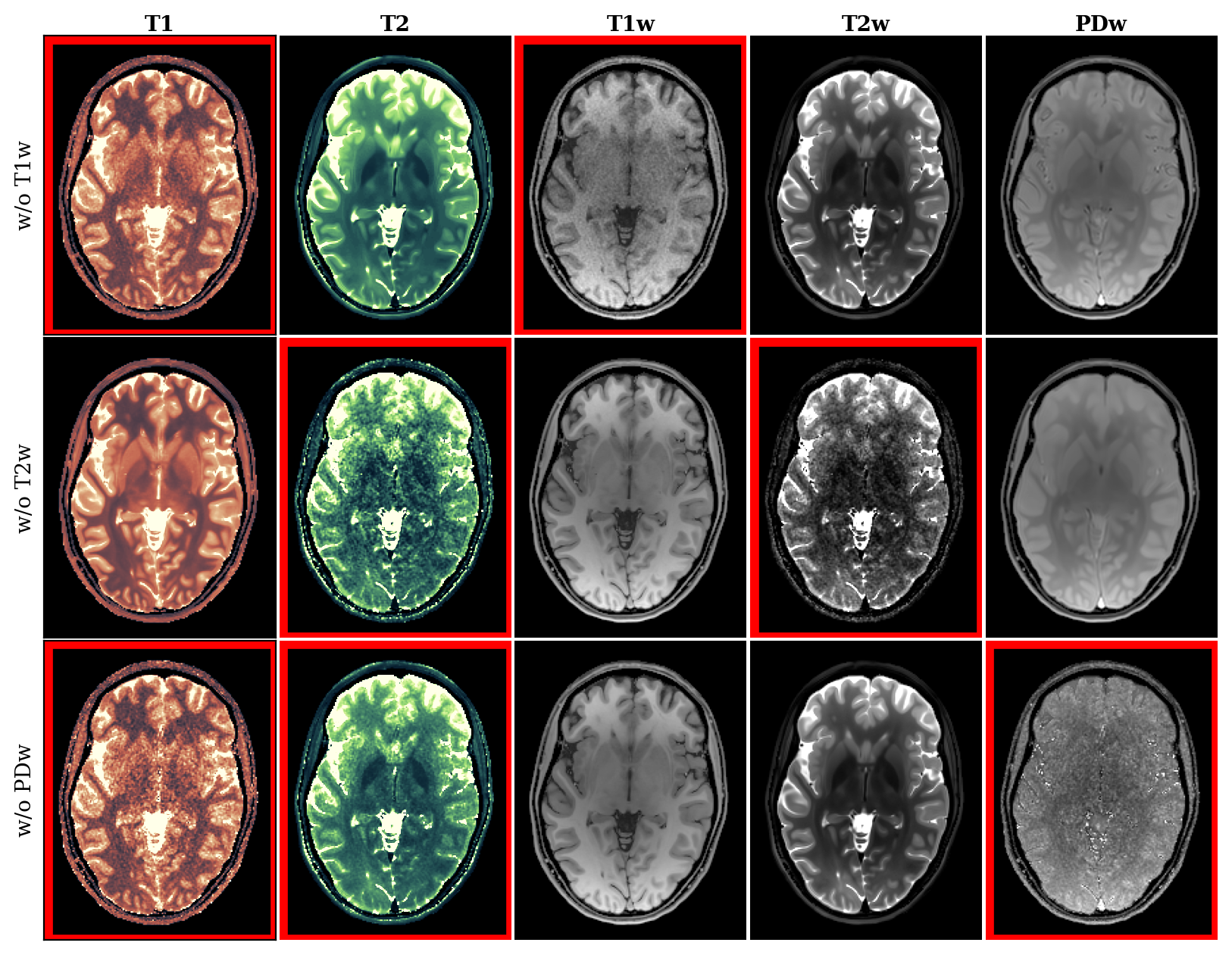}
    
\caption{T1/T2 maps and synthesized MRIs (columns) from \mriq\ with one denoising modality prior disabled (rows); most affected images in each row are highlighted. 
}
\label{fig:switchoffmodal_main}
\vspace{-.3cm}
\end{figure}

We assess the contribution of each MRI modality prior used in \mriq\ by disabling individual contrasts   (T1w, T2w, PDw) via setting their weights small 
$\overline{\gamma}_i=10^{-10}$, while other parameters unchanged. Table~\ref{tab:DNmodality-full} reports average T1/T2 mapping MAPEs and synthesized MRI PSNRs on the \invivo\ dataset, compared with the base model using all contrasts (Fig~\ref{fig:switchoffmodal_main} 
shows qualitative results). Using denoising priors from all three modalities gives the lowest combined T1+T2 error. 
% Excluding T1w or T2w priors primarily degrades the corresponding quantitative maps and contrast reconstructions. 
Excluding T1w priors primarily degrades T1 and T1w accuracies, while excluding T2w priors mostly  impacts T2 and T2w reconstructions. 
Removing the PDw prior increases both T1 and T2 errors, which is expected since MR signal evolutions depend jointly on proton density, and inaccurate PD estimation propagates to errors in T1 and T2 even when T1w and T2w priors are present. 

%%%%%%%%%%%%%%%%%%%%%%%%%%%%%%%%%
% ABLATION TABLE: Switch-off MRI modality with MAPES
%%%%%%%%%%%%%%%%%%%%%%%%%%%%%%%%%

%%%%%%%%%%%%%%%%%%%%%%%%%%%%%%%%%%% Table without STD 
%%%%%%%%%%%%%%%%%%%%%%%%%%%%%%%%%

\begin{table}[t!]
\scalebox{1.0}{
\setlength\extrarowheight{1pt}
    \centering
    \begin{tabular}{lccc|ccc}
    % \hline
    % \hline
    \toprule
         %\multirow{ 2}{*}
         {\textbf{\mriq}} 
         & \multicolumn{3}{c}{\textbf{MAPE} (\%) $\downarrow$} 
         & \multicolumn{3}{c}{\textbf{PSNR} (dB) $\uparrow$} 
         \\ \cline{2-7}
           Modalities  &\textbf{T1+T2} &\textbf{T1} & \textbf{T2}   
           &\textbf{T1w} &\textbf{T2w} & \textbf{PDw}
       
                    \\ 
                \hline 
        All (base) & 12.01 & 4.93 & 7.08  & 35.91 & 30.03 & 30.63  \\
        T2w PDw & 14.59 & {7.50} &  7.09 & {33.75 } & 29.94 & 30.66   \\
        T1w PDw & {16.51} & 4.77 & {11.74 } & 36.05 & {27.96} & 
        31.07\\
        T1w T2w & {19.19} & {10.44 } & {8.75} & 36.01 & 29.36 & {27.28} \\
        \bottomrule
  
    \end{tabular}
    }
     \caption{\mriq\ performance with all or partial MRI modality priors, averaged over \invivo\ data. 
     }
   \label{tab:DNmodality-full}
   \vspace{-.3cm}
    \end{table}

\subsubsection{MR synthesis parameters}
\label{sec:ablation_TRTE}
\begin{figure}[t!]
    \centering    \includegraphics[width=0.7\linewidth]{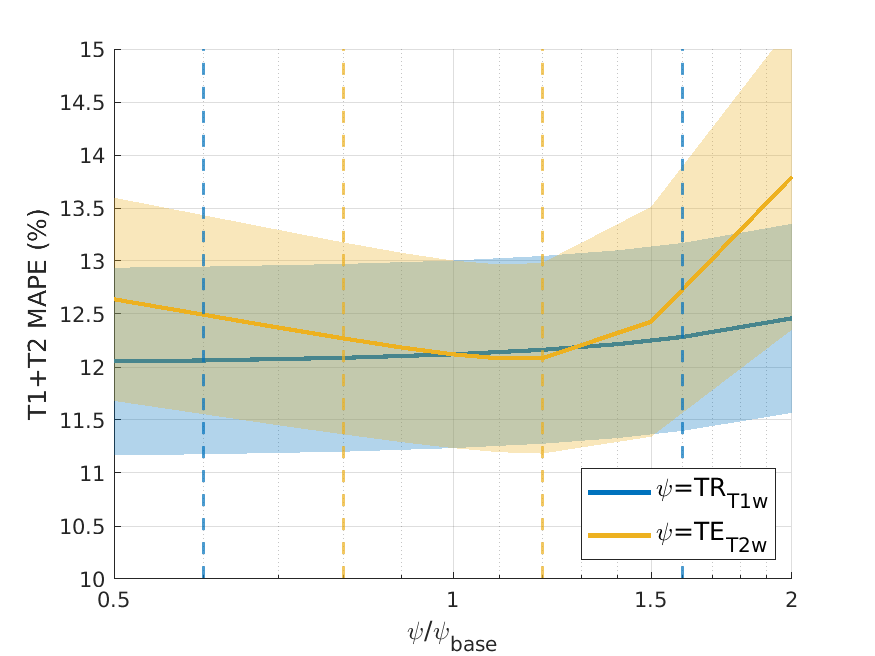}%
    \caption{ \mriq\ 
    mapping performance 
    % combined T1+T2 MAPE 
    on \invivo\ data vs. T1w TR (blue) and T2w TE (orange) variations relative to the default  synthesis values (TR$_\text{base}$/TE$_\text{base}$ = 500/100 ms). Solid lines and  shaded regions show  mean MAPE ± standard deviation. Dashed lines mark practical parameter ranges.}
\label{fig:mrparam_sensitivity}
\end{figure}

We examined the sensitivity of \mriq\ to the spin-echo synthesis parameters TR (T1w) and TE (T2w). The base model used TR $=500$~ms for T1w and TE $=100$~ms for T2w synthesis, chosen within typical clinical ranges (TR: 300–800 ms; TE: 80–120 ms). 
Robustness of T1/T2 mapping was evaluated on \invivo\ dataset by independently varying TR and TE from $0.5\times$  to $2\times$ their base values, with keeping other parameters fixed. As shown in Fig.~\ref{fig:mrparam_sensitivity}, reconstruction accuracy is slightly more sensitive to TE (T2w) than TR (T1w), yet remains overall stable, with combined T1+T2 MAPE varying by less than 0.2\% within clinically relevant ranges (dashed lines). 

\subsubsection{Runtime/memory comparison}
Table~\ref{tab:runtime_memory_baselines} reports reconstruction times and GPU memory usage of \mriq\ (base model) and selected baselines on \invivo\ data, and further decomposes the per-iteration cost of \mriq\ across its main components. The CG-based $\mathrm{prox}_f$ step dominates both runtime and memory, followed by MRI denoising and dictionary matching. ARNet achieves the fastest reconstruction, as it is non-iterative beyond CG initialization and does not enforce k-space consistency. Other methods 
using iterative forward–adjoint operations to 
minimize k-space loss incur longer but  overall comparable runtimes, with \mriq\ taking $\sim$11 minutes on a single GPU to reconstruct a 3D brain volume at 1~mm$^3$ resolution from multicoil non-Cartesian data. Studies below  further analyze runtime factors in \mriq. 

\begin{table}[t!]
  \centering
  \setlength{\tabcolsep}{2pt}
  \setlength\extrarowheight{1pt}
  \renewcommand{\arraystretch}{1.1}
  \begin{tabular}{lccccc|ccc}
    \toprule
     & \multicolumn{5}{c}{{Total cost}} & \multicolumn{3}{c}{{\mriq\ cost/iter.}} \\
    \cmidrule(lr){2-6} \cmidrule(lr){7-9}
     &\footnotesize{LLR}& \footnotesize{LRTV} & \footnotesize{MRFPnP} & \footnotesize{ARNet} & \footnotesize{\mriq}  & \footnotesize{Prox} & 
     \footnotesize{DM} & 
     \footnotesize{DN}   \\
    \midrule
    Runtime $\approx$   &   778 &528 &547 &89 &684 & 40.1 & 9.9 & 10.6 \\
    GPU Mem. & 33.9 & 33.9 & 36.9 &  34.8 & 36.9 & 36.9 & 19.5 & 26.5  \\
    \bottomrule
  \end{tabular}
  \caption{(Left) Reconstruction times (s) and peak GPU memory (GB) usage of each method, averaged over \invivo\ data. (Right) \mriq\ average per-iteration cost breakdown for $prox_f$, dictionary matching (DM), and denoising (DN). 
  %steps. Experiments used a single NVIDIA RTX A6000 GPU.
  }
\label{tab:runtime_memory_baselines}
\vspace{-.4cm}
\end{table}

\subsubsection{Annealed denoising schedule}
\label{sec:expe_ablation_anneal}
\begin{figure}[t!]
  \centering
\includegraphics[width=.9\linewidth]{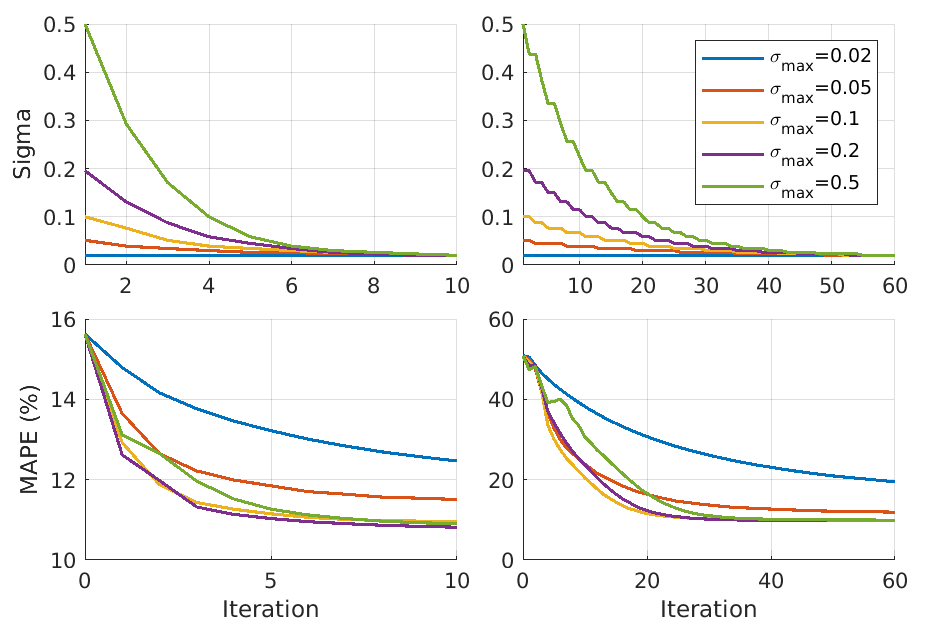}
  \caption{\mriq\ denoising schedules $\sigma_k$ (top) and T1+T2 MAPEs (bottom) vs. iterations for \invivo\ (left) and BrainWeb (right) reconstruction examples. 
  }
  \vspace{-.2cm}
  \label{fig:mape_iter}
\end{figure}

%===combinedvivo/BW====================
We compare fixed and annealed denoising schedules in terms of reconstruction speed and accuracy. Experiments varied the initial denoising level $\sigma_{\max}\in[0.02,0.5]$ and the number of iterations $K$, while keeping other parameters fixed. Denoising schedules were logarithmically spaced and rounded to discrete $\Sigma$ levels, with $\sigma_{\max}=\sigma_{\min}=0.02$ corresponding to a fixed schedule. 
Table~\ref{tab:bis_sigmax_k_combined} reports combined T1+T2 MAPE on representative \invivo\ and BrainWeb volumes (in Figs~\ref{fig:vivo-maps} and~S.1),
% ~\ref{fig:bw-maps}),
while Fig.~\ref{fig:mape_iter} shows the evolution of denoising levels $\sigma_k$ and reconstruction errors over iterations ($K=10/60$ for \invivo/BrainWeb). Fixed denoising schedules show the slowest progress on reducing errors. Annealed schedules ($\sigma_\text{max}>0.02$) enable faster reconstructions. 
For \invivo\ data, $\sigma_{\max}=0.2$ achieved the lowest error using $K=10$ iterations; larger $K$ were sub-optimal in runtime and accuracy. 
The same $\sigma_{\max}$ performed best on BrainWeb, although its single-coil setting required more iterations due to poorer conditioning.

%=================================================
% Figure: MAPE vs Iter v Sigma 
%=================================================

\begin{table}[t]
  \centering
  \scalebox{0.93}{
  \setlength{\tabcolsep}{3pt}
  \setlength\extrarowheight{3pt}
  \begin{tabular}{l *{5}{c}|c *{5}{c}}
    \toprule
    & \multicolumn{5}{c}{\emph{in-vivo} data} & \multicolumn{5}{c}{{BrainWeb data}} \\
    \cmidrule(lr){2-6} \cmidrule(lr){7-11} 
     $\sigma_{\max}$ & $K$= 5 & 10 & 15 & 20 & 30 & 10 & 20 & 40 & 60 & 100 \\
    \midrule
    %\multirow{5}{*}{\rotatebox{90}{\emph{In-vivo}}}
     0.02 & 13.22 & 12.46 & 12.10 & 11.87 & 11.65 
    %&\multirow{5}{*}{\rotatebox{90}{\emph{In-vivo}}} 
    & 38.19 & 30.72 & 23.07 & 19.50 & 16.46\\
     0.05 & 12.17 & 11.51 & 11.19 & 11.11 & 11.12 
    & 32.51 & 23.43 & 15.13 & 11.91 & 10.25\\
     0.1  & 11.54 & 10.95 & 10.88 & 10.97 & 11.23 
    & 29.40 & 19.46 & 11.61 & 9.90 & 9.47\\
     0.2  & 11.27 & 10.82 & 10.91 & 11.11 & 11.35 
    & 28.59 & 18.87 & 11.02 & 9.61 & 9.45\\
     0.5  & 11.34 & 10.91 & 11.01 & 11.20 & 11.46 
    & 29.12 & 19.37 & 11.31 & 9.93 & 9.49\\
    \bottomrule
  \end{tabular}
  }
  \caption{Combined T1+T2 MAPE (\%) across $K$, $\sigma_{\max}$ on \invivo/BrainWeb reconstruction examples.
  % \mriq\ combined T1+T2 MAPE (\%) across combinations of $K$ and $\sigma_{\max}$ on representative volumes from \invivo\ and BrainWeb datasets.
  }
  \label{tab:bis_sigmax_k_combined}
  \vspace{-.2cm}
\end{table}

\begin{table}[t!]
\centering
\setlength{\tabcolsep}{4pt}
\setlength\extrarowheight{1pt}
\begin{tabular}{lccccc}
\toprule
$\kcg$ & 1 & 5 & 10 & 20 & 40 \\
\midrule
MAPE (\%)        & 12.53 & 10.88 & 10.82 & 10.81 & 10.81 \\
Avg./iter (s)   & 5.8   & 23.0  & 39.9            & 61.3  & 87.0  \\
\bottomrule
\end{tabular}
\caption{\mriq\ T1+T2 MAPE (\%) and $prox_f$ average runtime (s) vs. CG iterations ($\kcg$) on an \invivo\ volume.}
\label{tab:vivo_cg}
\vspace{-.45cm}
\end{table}

\subsubsection{Accuracy of CG updates}
\label{sec:ablation_CG}
We analyze runtime–accuracy trade-off of $\kcg$, the maximum number of CG iterations in the k-space-enforcing $prox_f$ updates (Algorithm~\ref{alg:mri2qmap}, Line~10). Table~\ref{tab:vivo_cg} reports average per-$prox_f$ runtime and T1+T2 MAPE errors on a representative \invivo\ volume, with CG terminating at a relative residual of $10^{-5}$ or at $\kcg$. $\kcg\in[5,10]$ provide a favorable trade-off: smaller values increased error, while larger ones yield only marginal gains ($\leq$0.01\%) at higher runtime cost.  
Trends are similar on BrainWeb data (Supplementary Table~S.1), 
overall indicating that depending on problem conditioning, CG updates require some moderate, but not high-precision convergence.

\subsubsection{Inexact dictionary matching}
\label{sec:ablation_gdm}
Dictionary matching steps can be  accelerated using approximate search methods~\cite{cauley2015fastgdm,golbabaee2019coverblip}. We evaluate Fast Group Matching (FGM), which clusters dictionary atoms by cosine similarity and restricts matching to a small subset of relevant groups, as detailed in~\cite{cauley2015fastgdm}. In our experiments, FGM used 100 groups with cosine similarity thresholds of 0.95 (in vivo) and 0.98 (BrainWeb), chosen to balance accuracy and speed. Using FGM within \mriq\ produced nearly identical reconstructions (accuracy loss $\leq$0.07\% in combined T1+T2 MAPE), while achieving a $4-5.5\times$ speedup in dictionary matching  steps ($\sim$1.7s for BrainWeb and $\sim$2.4s for \invivo\ data, 
%with more complex bio-parameter distributions, 
down from $\sim$10s for exact matching).

\subsubsection{Other acquisition settings}
Finally, Table~\ref{tab:supp_R2R8}
provides reconstruction results (zero-shot, without retraining the denoiser model) for \invivo\ data on other acceleration factors $R=2,4,6$, showing consistent performance of \mriq\ across different acquisition settings. 
% As with $R=8$, the  $R=2, 4, 6$  acquisitions were obtained by sub-sampling data from the gold-standard scans to retain 24, 12, and 8 readout groups, respectively, followed by the same pre-processing steps.  
Results are compared with representative compressed-sensing and deep learning methods (LRTV, MRF-PnP); 
ARNet is excluded, since it  requires retraining for each $R$. 
As $R$ increases, errors rise for all methods, but deep learning approaches (\mriq, MRF-PnP) tend to remain more robust. \mriq\ achieves metrics comparable to supervised MRF-PnP despite no MRF training data (for qualitative comparisons see Fig~\ref{fig:R2R8_vivo}).

%%%%%%%%%%%%%%%%%%%%%%%%%%%%%%%%%
% ABLATION TABLE: R2 R4 R6 
%%%%%%%%%%%%%%%%%%%%%%%%%%%%%%%%%
% \newpage
\begin{table}[t!]
\scalebox{1.0}{
    \centering
    \setlength{\tabcolsep}{5pt}
     \setlength\extrarowheight{1pt}
    \begin{tabular}{llccc|ccc}
    % \hline
    % \hline
    \toprule
         &\multirow{2}{*}
         {\textbf{Method}} 
         & \multicolumn{3}{c}{\textbf{MAPE} (\%) $\downarrow$} 
         & \multicolumn{3}{c}{\textbf{PSNR} (dB) $\uparrow$} 
         \\ 
         \cline{3-8}
         &&\textbf{T1+T2} &\textbf{T1} &\textbf{T2}   &\textbf{T1w} &\textbf{T2w} & \textbf{PDw}\\ 
         \hline 
        \multirow{3}{*}{\rotatebox{90}{$R=2$}} & LRTV & 7.94 & 3.45 & 4.49 & 40.32 & 34.80 & 35.30   \\
        &MRFPnP & 7.56 & 3.15& 4.41& 41.66&  35.74& 36.23 \\
         &\textbf{\mriq} & 7.50 & 3.25 &4.25 & 41.58& 35.44& 35.13 \\
         \midrule
         \multirow{3}{*}{\rotatebox{90}{$R=4$}} & LRTV & 11.01 & 4.88 & 6.13 & 35.88 &  31.18 & 32.54\\
        &MRFPnP & 10.07 & 4.20& 5.87& 38.62&  32.85& 34.02  \\
         &\textbf{\mriq} & 9.88 & 4.18& 5.70& 38.39& 32.27& 32.87  \\
         \midrule
         \multirow{3}{*}{\rotatebox{90}{$R=6$}} & LRTV & 13.05 & 6.03& 7.02& 34.06 & 29.58& 31.24\\
        &MRFPnP & 11.83 & 5.10 & 6.73& 36.58& 30.82& 32.52 \\
         &\textbf{\mriq} & 11.35 &4.74& 6.61& 36.67& 30.70& 31.51 \\
         
        \bottomrule
  
    \end{tabular}
    }
     \caption{\mriq\ performance vs. selected baselines on acceleration factors $R=2,4,6$, averaged on \invivo\ dataset. 
     % $(^*)$ indicates MRF-trained methods.
     }
     \vspace{-.5cm}
\label{tab:supp_R2R8}
    \end{table}

\vspace{-.1cm}
\subsection{Limitations and future work}
While the proposed framework shows promising results for MRF reconstruction, several extensions could further broaden its applicability. The current model reconstructs single-compartment, multi-parametric maps; future work could incorporate multi-compartment models to account for partial volume effects~\cite{bayesianmrf_mcgiveney2018}. 
Motion artifacts are relevant in clinical applications and require the integration of motion-compensation modules~\cite{cruz2022mrf-nonrigidmotion}. 
Consistent with the preprocessing pipeline in~\cite{delics2025deep}, B0/B1 corrections were omitted; but these could be complementarily added to further improve quantitative accuracy~\cite{cao2022tgas}. Extensions to other qMRI sequences and bio-parameters are also natural directions.
From a computational perspective, further speedups are possible through stochastic optimization~\cite{mayo2024stodip}, improved preconditioning   
% of data-fidelity term 
and multi-GPU parallelization~\cite{iyer2024polynomialprecond}.  
Incorporating additional MRI contrasts (e.g. FLAIR) to further enrich multimodal priors, adopting more expressive denoising backbones~\cite{swin_mridenoiser_mrm2024}, and possible integration with denoising diffusion models for probabilistic reconstruction~\cite{eucker2023_ddpm_mri} remain avenues for future investigation.

% ==============================================================
% = C O N C L U S I O N S
% ==============================================================
% \vspace{-.1cm}
\section{Conclusions}
\label{sec:conclusions}
We show that image priors learned from independently acquired routine weighted-MRI datasets can support quantitative MRI reconstruction. We introduced MRI2Qmap, a plug-and-play framework based on deep denoising autoencoders pretrained on large multimodal MRI datasets. The method was validated on accelerated 3D whole-brain MRF data from \emph{in vivo} and simulated acquisitions, achieving competitive or improved performance relative to MRF baselines without requiring ground-truth quantitative training data. Given the broad availability of routine MRI, this approach may enable more scalable, data-driven quantitative reconstruction using larger and more diverse datasets.
\vspace{-.1cm}

% ---- Bibliography ----
% \bibliographystyle{splncs04}
\bibliographystyle{IEEEtran}
\bibliography{references}

%%%%%%%%%% Merge with supplemental materials %%%%%%%%%%
% \input{sections/supplementary}
% \appendices
% =====================================
% SUPPLEMENTARY MATERIALS
% ======================================

\clearpage
\newpage

% Restart numbering logic
\setcounter{section}{0}
\setcounter{equation}{0}
\setcounter{figure}{0}
\setcounter{table}{0}
\setcounter{page}{1} % Optional: restart page numbering for the supplement

% Redefine formats to include "S"
\renewcommand{\thesection}{S.\Roman{section}}
\renewcommand{\theequation}{S.\arabic{equation}}
\renewcommand{\thefigure}{S.\arabic{figure}}
\renewcommand{\thetable}{S.\arabic{table}}

\section*{MRI2Qmap: Supplementary materials}

\section{Proof of expressions~(10) and (11)}
\label{sec:app_dm_proof}
Dictionary matching is performed independently at each voxel; for brevity, we omit the voxel index.
Let $\tilde \x \in \Cbb^s$ and $\tilde \m \in \Rbb_+^{s'}$ denote the signals of a voxel from a $s$-channel TSMI and a stack of $s'$-channel MRI modalities, respectively. Let $\Dq_j\in \Cbb^{s}$ and $\Dw_j\in \Rbb_+^{s'}$ respectively denote atoms of the MRF and MRI dictionaries indexed by $j\in\textsc{LUT}$ indicating a certain tissue property. The combined {MRF-MRI dictionary matching is defined as:
\begin{align}
\argmin_{j\in \textsc{LUT},\PD\in \Cbb} \|\tilde \x - \PD . \Dq_j \|^2_2 + \left\| \tilde \m - |\PD|. \Dw_j \right\|_{\bm{\gamma}}^2, 
\end{align}
or equivalently, 
\begin{align}
\label{eq:cdm_}
\argmin_{j\in \textsc{LUT},\PD\in \Cbb} |\PD|^2 A_j -2 \mathrm{Re}\langle \tilde \x, \PD \Dq_j \rangle -2 \langle \tilde \m, |\PD|\Dw_j \rangle_{\bm{\gamma}}
\end{align}
where $A_j := \|\Dq_j\|^2_2 +\|\Dw_j\|_{\bm{\gamma}}^2$. 
The objective of \eqref{eq:cdm_} depends on the PD's phase (denoted by $\angle$) only in its second term, and is minimized by setting the optimal phase $\angle \PD^* = -\angle \langle \x,\Dq_j \rangle$ or equivalently:
\begin{align}
\label{eq:pdphase}
\angle \PD^* = \angle \langle \Dq_j, \tilde \x \rangle = \frac{\langle \Dq_j, \tilde \x \rangle}{|\langle \Dq_j,\tilde \x \rangle|}.
\end{align}
Substituting~\eqref{eq:pdphase} in the objective of \eqref{eq:cdm_} gives: 
% this optimal PD phase, gives the objective of \eqref{eq:cdm_} as follows: 
\begin{align}
\label{eq:cdm__}
|\PD|^2 A_j - 2|\PD| B_j,
\end{align}
where $B_j := |\langle \tilde \x,\Dq_j \rangle| + \langle \tilde \m,\Dw_j \rangle_{\bm{\gamma}}>0$. The minimizing PD magnitude is therefore: 
\begin{align}
\label{eq:pdmag}
|\PD^*|=B_j/A_j.
\end{align}
Together, the optimal phase and magnitude prove the closed-form expression~(11). %\eqref{eq:optpd}. 
%Equations~\ref{eq:pdmag} and \ref{eq:pdphase} together prove the expression~\eqref{eq:optpd}
Substituting \eqref{eq:pdmag} in \eqref{eq:cdm__} yields $ -B_j^2/A_j$ as  for the objective of \eqref{eq:cdm_}, 
which is minimized by searching over LUT the $j^*$ that maximizes $B_j/\sqrt{A_j}$ as in (10), %\eqref{eq:fused-DM-search}, 
which completes the proof.

\section{Simulated brainweb dataset}
\label{app:bw}
The simulated dataset was generated using 3D quantitative T1, T2, and PD maps produced with the MRtwin software~\cite{MRtwin2025} from anatomical brain models in the BrainWeb database~[48]. %\cite{brainweb}. 
BrainWeb provides 20 normal-brain models at 0.5 mm isotropic resolution; these were resampled to 1 mm$^3$ isotropic resolution and cropped to a $256^3$-voxel volume. MRtwin constructed quantitative maps using BrainWeb’s fuzzy tissue-segmentation maps and their assigned $(T_1\text{ ms}, T_2\text{ ms}, \text{PD au})$ properties for the following tissue classes: Background/skull (0, 0, 0), Cerebrospinal fluid (4000, 1990, 1), Gray matter (1250, 90, 85), White matter (750, 70, 75), Fat (250, 70, 1), Muscle/skin (1485, 50, 1), Vessels (1960, 200, 1.5), Around fat (250, 70, 1), Dura matter (990, 150, 1), Bone marrow (587, 50, 1). For evaluation, binary tissue masks were derived from the ground-truth fuzzy segmentation maps by thresholding tissue probabilities at 90\%.

\section{Implementation details}
\label{app:denoiser_IN}

\subsubsection*{Baselines methods}
\label{app:baselines}

\textbf{LLR}~[5] %\cite{cao2022tgas} 
and \textbf{LRTV}~[8] 
%\cite{golbabaee2021lrtv} 
are convex reconstruction methods that enforce k-space consistency with TSMI regularization based on locally low-rank and total variation priors, respectively. The regularization strength parameters $\lambda$ was set to $\lambda=$ 3e-6$/$5e-7 for LLR and $\lambda=$ 2e-7$/$4e-8 for LRTV in \invivo$/$BrainWeb experiments.
Both methods employ FISTA, using 60 iterations for \invivo\ data (multicoil, better conditioned, faster convergence) and 500 iterations for single-coil BrainWeb data (slower convergence).
For LRTV, $prox_\Rcal$ used 
%PyTorch implementation of 
the Chambolle-Pock algorithm, while LLR employed singular-value soft-thresholding over non-overlapping TSMI patches (spatial size $8^3$) from the toolbox in~[11].  %~\cite{delics2025deep}. 
GPU acceleration was used for both proximal operators.

\textbf{ARNet} used the same U-Net backbone as \mriq\ denoiser, with input/output channel dimensions modified for TSMI restoration (10 channels for both input and output, representing the stack of real and imaginary components of the $s=5$ complex-valued TSMI channels). The class- and noise-level conditioning 
% mechanisms used in \mriq\ 
were not applicable and therefore disabled. Training used paired MRF data. Inputs ($\x_{\text{init}}$) were CG-reconstructed TSMI from $R=8$ accelerated acquisitions (the same initialization used in \mriq\ reconstruction), 
%except for the Brainweb simulated data, where MRF-ADMM was used for superior input quality).
and targets were reference TSMIs generated either from ground-truth qmaps in BrainWeb experiment, or from qmaps obtained from the gold-standard $R=1$ acquisitions for \invivo\ dataset. All TSMI volumes were normalized to $[-1,1]$. The network was trained using MSE loss on randomly cropped input–target patches (spatial size $64^3$), with random intensity scaling (up to $\pm 50\%$) for augmentation, using Adam optimizer, learning rate $10^{-4}$, batch size 5, and 10k training epochs. During inference, ARNet was used to restore CG-reconstructed TSMI patches (overlapping patches of stride size 32 was used).

\textbf{MRF-PnP}~[29] 
%~\cite{fatania2022plug} 
is an ADMM-based plug-and-play algorithm for TSMI reconstruction $\min_{\x} f(\x)+\Rcal(\x)$ based on enforcing k-space data fidelity and a data-driven TSMI Gaussian DAE prior replacing the regularizer’s proximal step. Unlike \mriq, the DAE directly restores MRF timeframe data;  training therefore required high-quality TSMIs, obtained from ground-truth qmaps (BrainWeb experiment) or from long, densely sampled $R=1$ MRF acquisitions (\invivo\ experiment). 
The TSMI denoiser employed the same UNet architecture as \mriq, with 10-channel TSMI input/output formed by stacking real and imaginary components. For training, TSMIs were normalized to $[-1,1]$, randomly cropped into patches (spatial size $64^3$), random intensity scaling (up to $\pm 50\%$), and corrupted with additive Gaussian noise at experimentally selected standard deviations $\sigma \in \{0.1, 0.05, 0.025, 0.01\}$ using noise-level conditioning; class conditioning was disabled. Models were trained with MSE loss and Adam optimization (learning rate $10^{-4}$, batch size 5) for 10k epochs. Reconstruction from $R=8$ acquisitions was initialized with CG-reconstructed TSMI $\x_{\text{init}}$ as in \mriq, and alternated CG-based $ \mathrm{prox}_f $ updates with TSMI denoising (random-shifts augmentation and non-overlapped patch denoising, like \mriq). Reconstruction parameters were $K=\kcg=10$, $\sigma=0.025$, $\mu=10^{-5}$ for \invivo\ data and $K=60$, $\kcg=20$, $\sigma=0.05$, $\mu=5\times10^{-6}$ for BrainWeb, where $K$ and $\kcg$ denote the PnP and inner CG iterations, respectively. The denoising level  $\sigma$ and the ADMM penalty $\mu$ parameters were kept fixed during iterations following~[29] 
% \cite{fatania2022plug} 
with values searched to yield best quantitative mapping performance.

\section{Additional  experimental results}
\label{app:expe}

\subsection{Reconstructed maps from BrainWeb experiment} 
The supplementary Fig.~\ref{fig:bw-maps} presents reconstructed T1/T2 maps from \mriq\ and baseline methods for a representative BrainWeb test volume at an acquisition acceleration factor of $R=8$, together with corresponding percentage error maps relative to the ground-truth reference. 
The MRF-ADMM baseline without spatial regularization exhibits pronounced aliasing artifacts. Incorporating spatial regularization via LLR/LRTV improves reconstruction quality; however, these methods generally incur higher errors than deep learning baselines (ARNet, MRF-PnP, and \mriq) due to residual aliasing and/or over-smoothing. \mriq\ shows the lowest errors, comparable to the best-performing supervised MRF-domain baseline (MRF-PnP), despite not being trained on MRF data.

\subsection{Reconstructed maps from other in-vivo subjects} 

The supplementary Figs.~\ref{fig:supp_vols_ax} and~\ref{fig:supp_vols_cor} show the axial and coronal views, respectively, of the reconstructed \invivo\ T1/T2 maps at acceleration factor $R=8$. 
Results are shown for \mriq\ and 
% selected 
representative compressed-sensing and deep-learning 
baselines (LRTV and ARNet) for the remaining \invivo\ subjects not included in the main manuscript figures. 
% Figs.~\ref{fig:vivo-maps} and~\ref{fig:mri2q_vols}.

\subsection{Accuracy of CG updates: BrainWeb experiment}
\label{sec:ablation_CG}
To complement the \invivo\ analysis in Table~V, 
% ~\ref{tab:vivo_cg}, 
we provide the runtime–accuracy analysis of the maximum number of CG iterations ($\kcg$) in the $prox_f$ updates for simulated single-coil BrainWeb reconstructions. Table~\ref{tab:cg_bw} reports average per-$prox_f$ runtime and the final T1+T2 MAPE reconstruction errors for a representative BrainWeb volume (in Fig.~\ref{fig:bw-maps}), with CG stopping at a relative residual of $10^{-5}$ or at the iteration limit  $\kcg$. The base-model choice $\kcg=20$ for BrainWeb experiments  provides a favorable trade-off in runtime-accuracy: smaller $\kcg$ increases errors, while larger values give relatively small accuracy gains despite  higher runtime cost. We also note that poorer conditioning of single-coil reconstructions necessitates larger $\kcg$ in the $prox_f$ step for BrainWeb than for \invivo\ data, although each CG iteration is faster than in multicoil $\invivo$ reconstructions.  

%%%%%%%%%%%%%%%%%%%%%%
% Table: Ablation Kcg (BW)
%%%%%%%%%%%%%%%%%%%%%%
\begin{table}[H]
\centering
\label{tab:proxf_invivo_horizontal}
% \setlength{\tabcolsep}{6pt}
% \renewcommand{\arraystretch}{1.1}
% \footnotesize
\begin{tabular}{lccccc}
\toprule
$\kcg$ & 5 & 10 & 20 & 40 & 60 \\
\midrule
MAPE (\%)        & 20.18 & 11.67 & 9.61 & 9.49 & 9.49 \\
Avg./iter (s)   & 2.5  & 4.5 & 8.4 & 14.4  & 18.1  \\
\bottomrule
\end{tabular}
\caption{\mriq\ T1+T2 MAPE (\%) and $prox_f$ average runtime (s) vs. CG iterations ($\kcg$) on BrainWeb data.}
\label{tab:cg_bw}
\end{table}

%%%%%%%%%%%%%%%%%%%%%%%%%%%%%%%%%%%%
% MRI2Q vs Baselines (BrainWeb)
%%%%%%%%%%%%%%%%%%%%%%%%%%%%%%%%%%%%
\begin{figure*}[t!]
    \centering
    \includegraphics[width=0.95\linewidth]{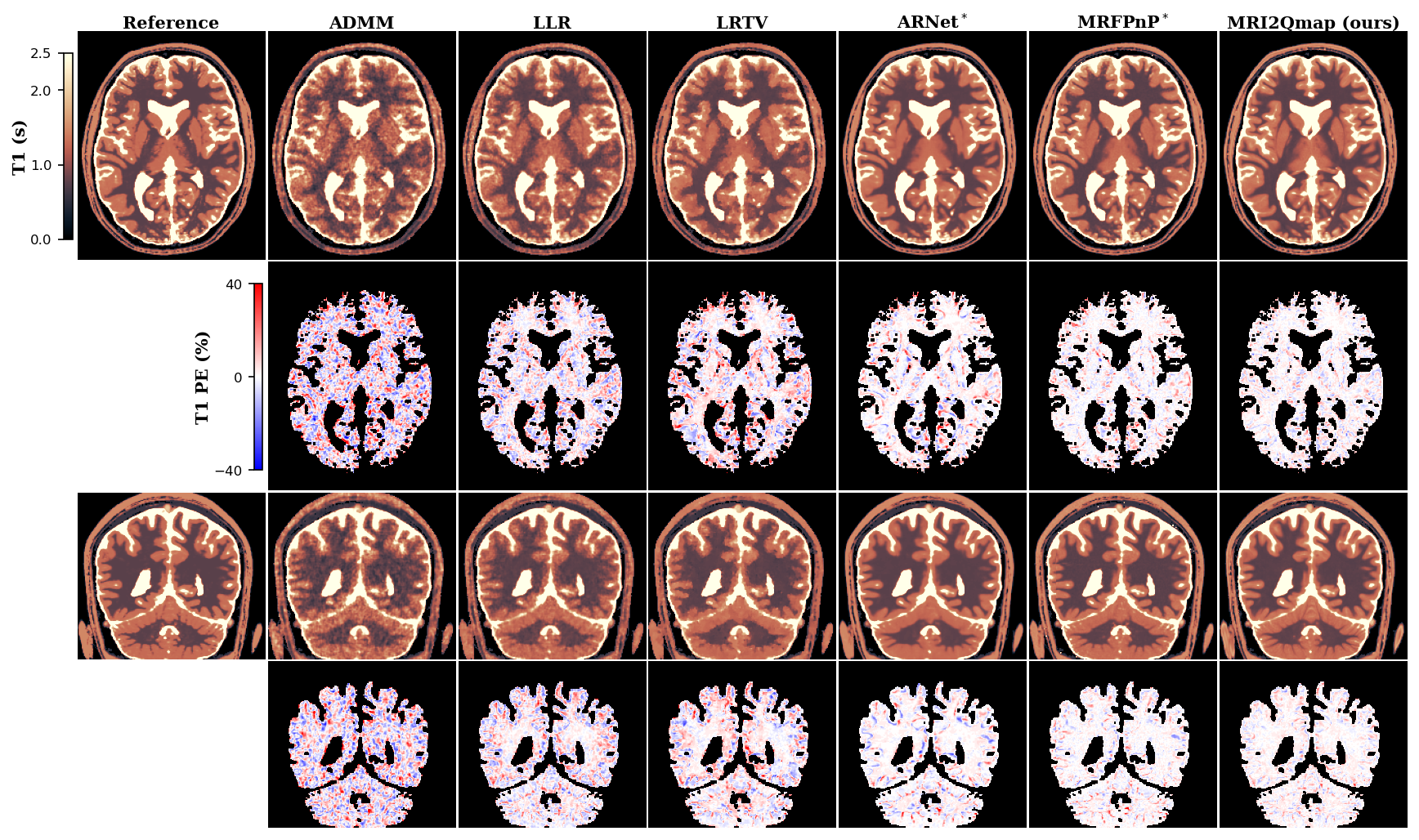}\\
    \includegraphics[width=0.95\linewidth]{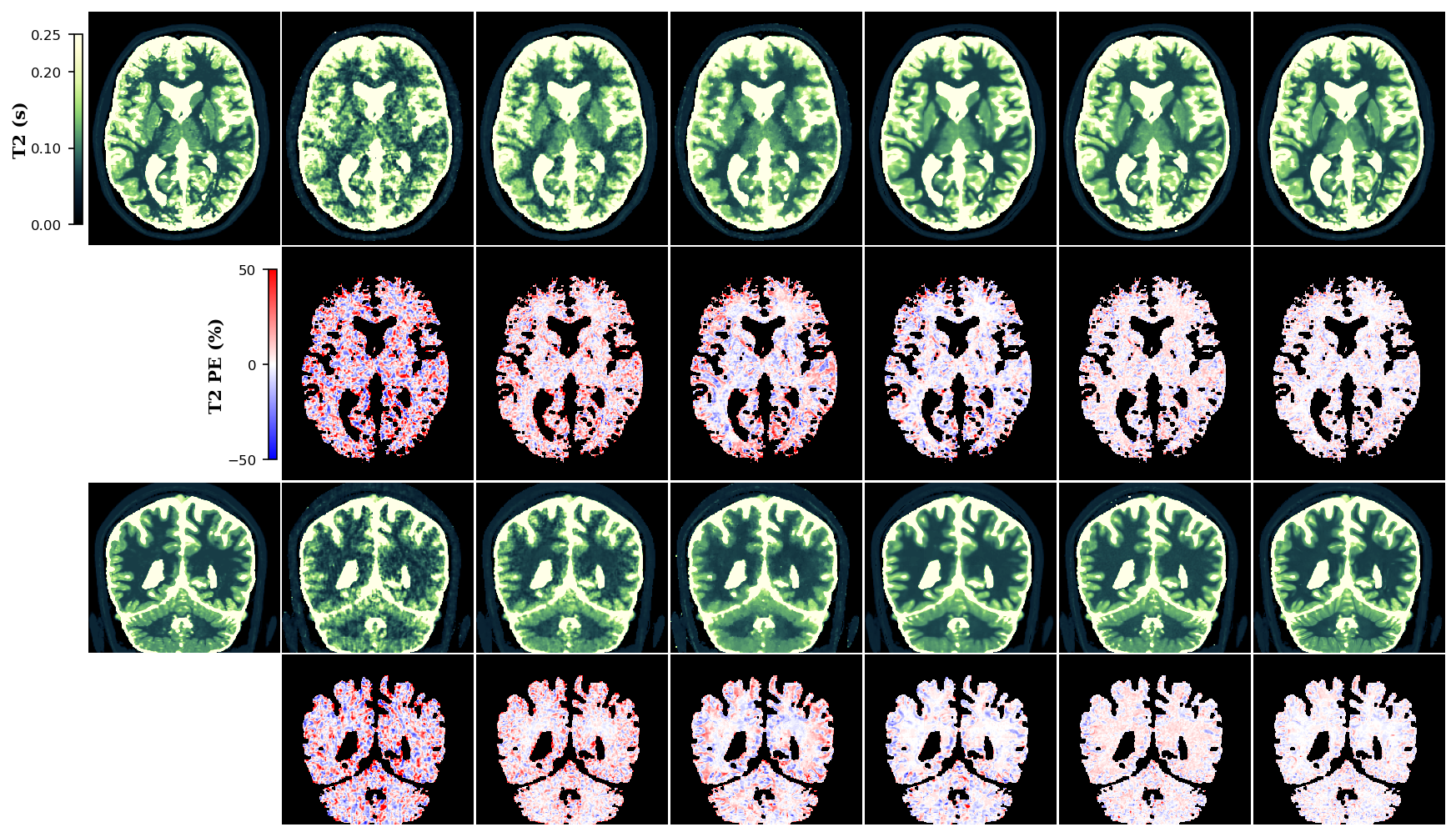}%
    \caption{
   Reconstructed T1/T2 maps (axial and coronal views) at acceleration factor $R=8$ for a representative simulated BrainWeb subject, comparing \mriq\ with MRF baselines ($^*$ indicates MRF-trained methods). Percentage error (PE) maps relative to the ground-truth reference are shown below each reconstruction.
    }
\label{fig:bw-maps}
\end{figure*}

\begin{figure*}[ht]
    \centering
    \begin{minipage}[T]{0.49\linewidth}
        \centering
        \includegraphics[width=\linewidth]{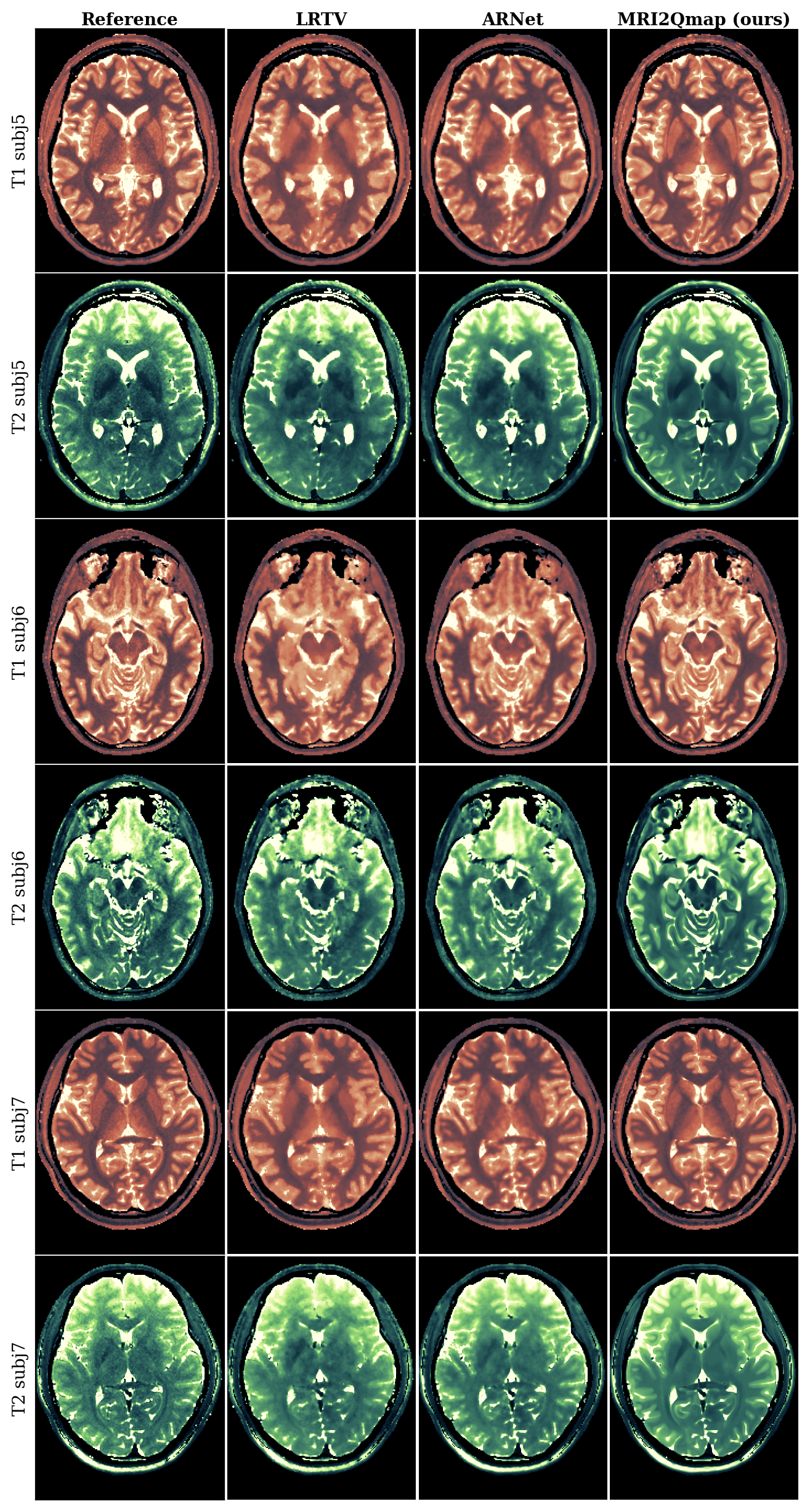} 
        \end{minipage} 
        \begin{minipage}[T]{0.49\linewidth}
        \centering
        \includegraphics[width=\linewidth]{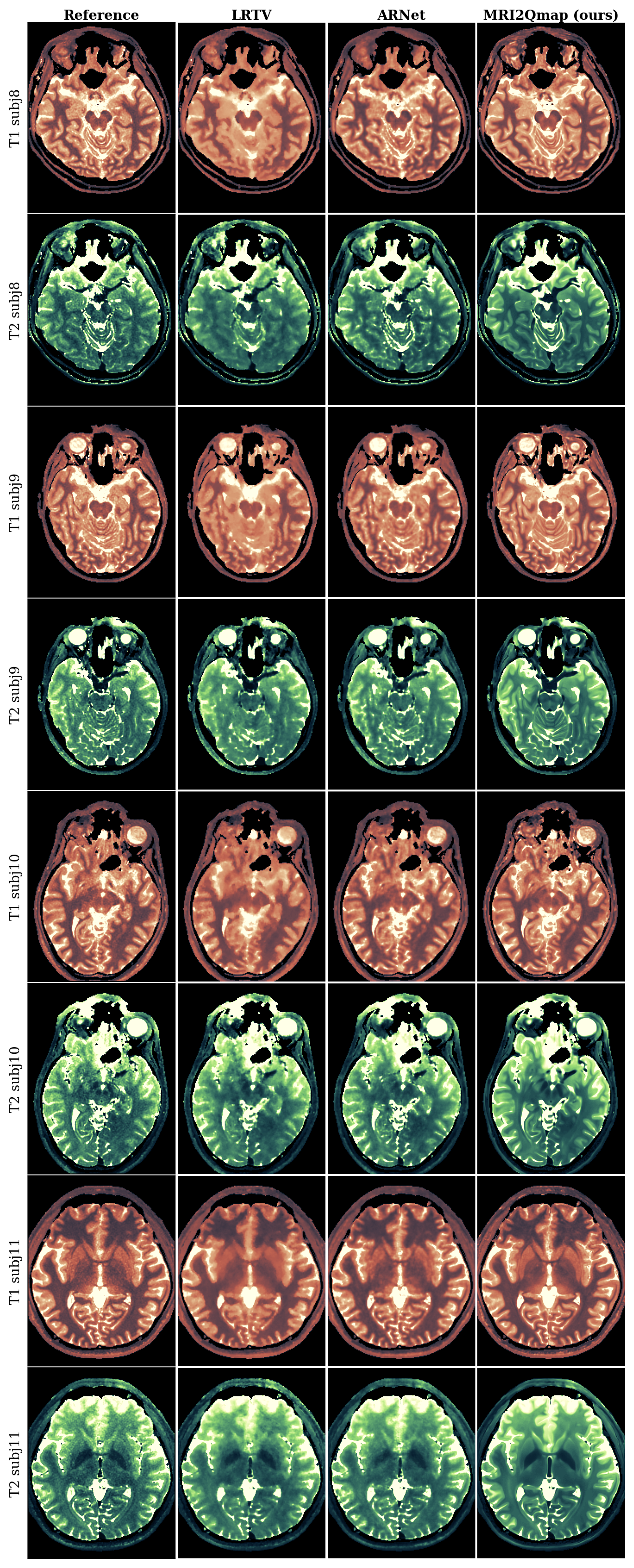}
        \end{minipage}
\caption{Reconstructed T1 and T2 maps (axial view) at $R=8$ for the remaining seven subjects in \invivo\ dataset, comparing \mriq\ with representative baselines (LRTV and ARNet). Relative to baselines, \mriq\ exhibits reduced artifacts and improved anatomical delineation (electronic zooming recommended).}
\label{fig:supp_vols_ax}

\end{figure*}
\begin{figure*}[ht]
    \centering
    \begin{minipage}[T]{0.49\linewidth}
        \centering
        \includegraphics[width=\linewidth]{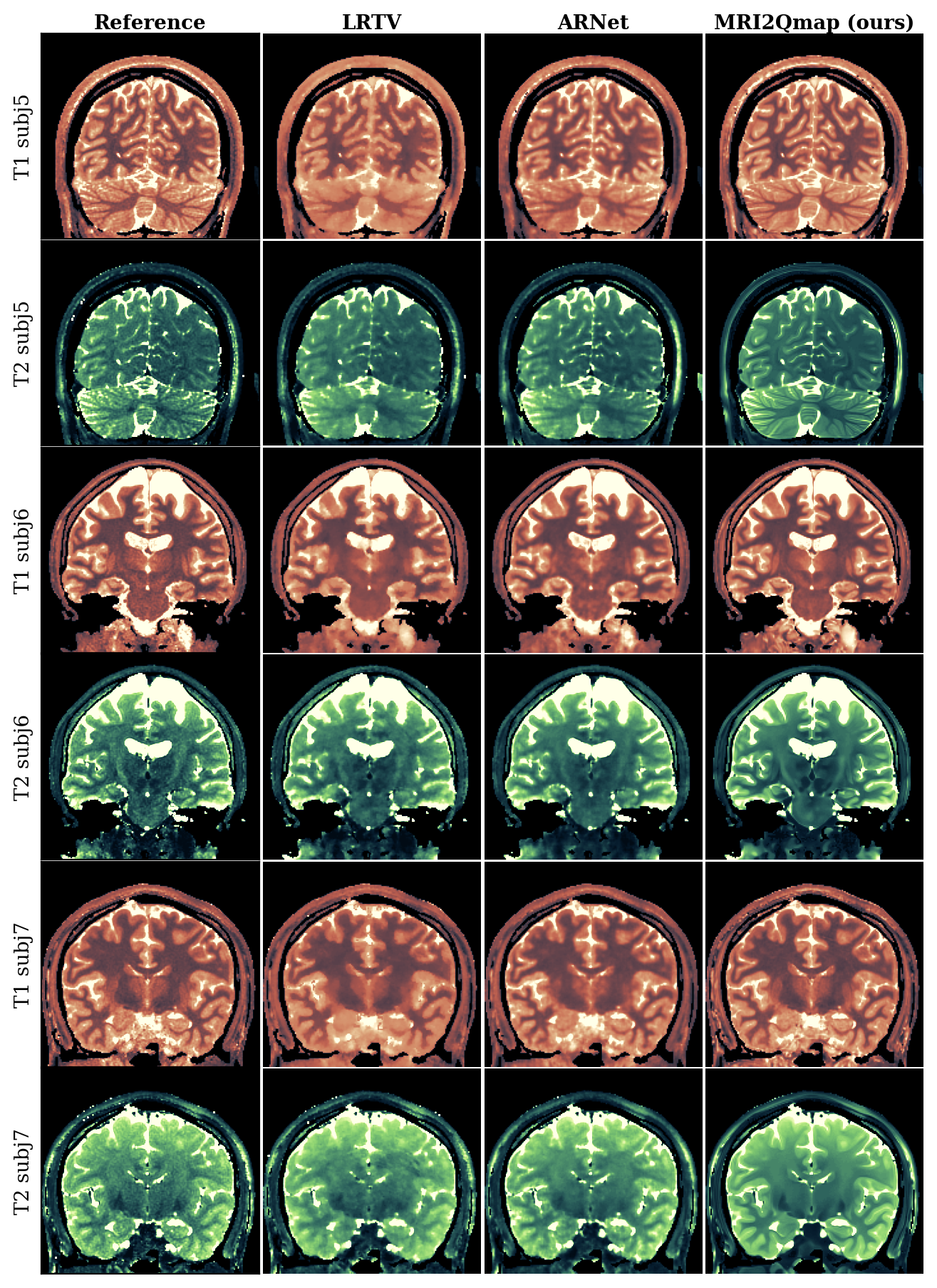} 
        \end{minipage} 
        \begin{minipage}[T]{0.49\linewidth}
        \centering
        \includegraphics[width=\linewidth]{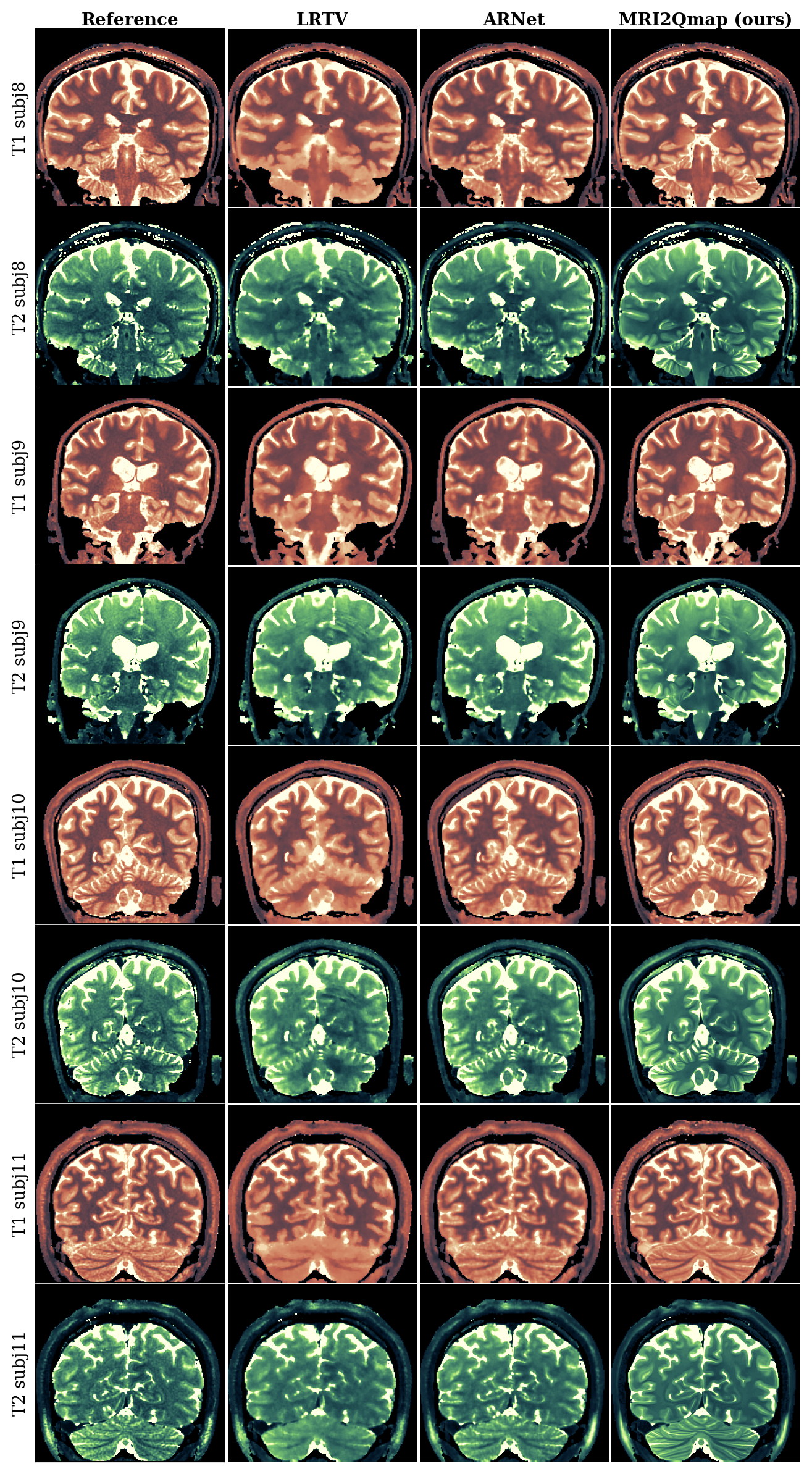}
        \end{minipage}
\caption{Reconstructed T1 and T2 maps (coronal view) at $R=8$ for the remaining seven subjects in \invivo\ dataset, comparing \mriq\ with representative baselines (LRTV and ARNet). Relative to baselines, \mriq\ exhibits reduced artifacts and improved anatomical delineation (electronic zooming recommended).}
\label{fig:supp_vols_cor}
\end{figure*}

%%%%%%%%%%%%%%%%%%%%%%
% Fig Ablation R2 to R8
%%%%%%%%%%%%%%%%%%%%%%

\begin{figure*}[t!]
    \centering
\includegraphics[width=.49\linewidth]{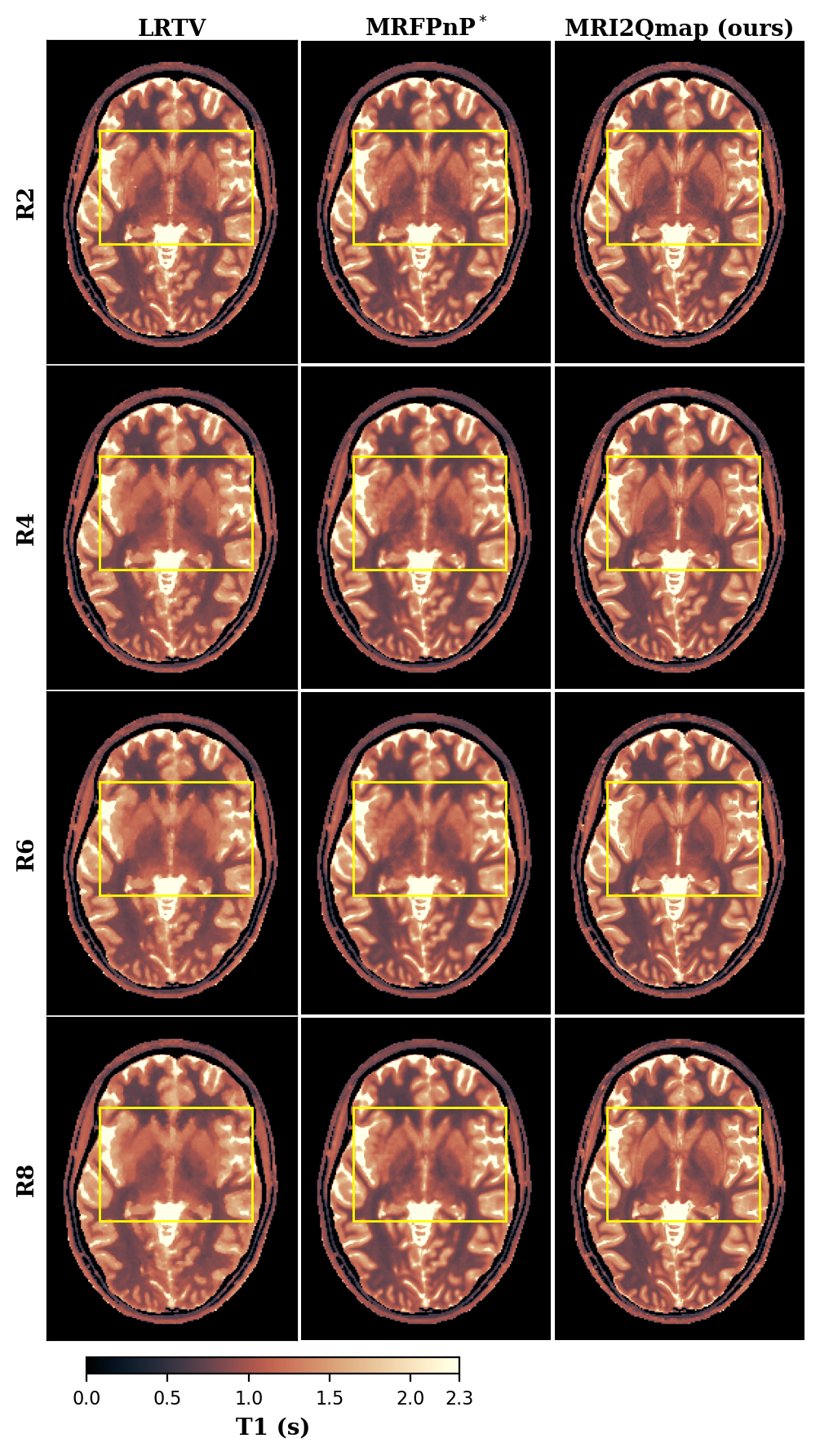} \hfill
\includegraphics[width=.49\linewidth]{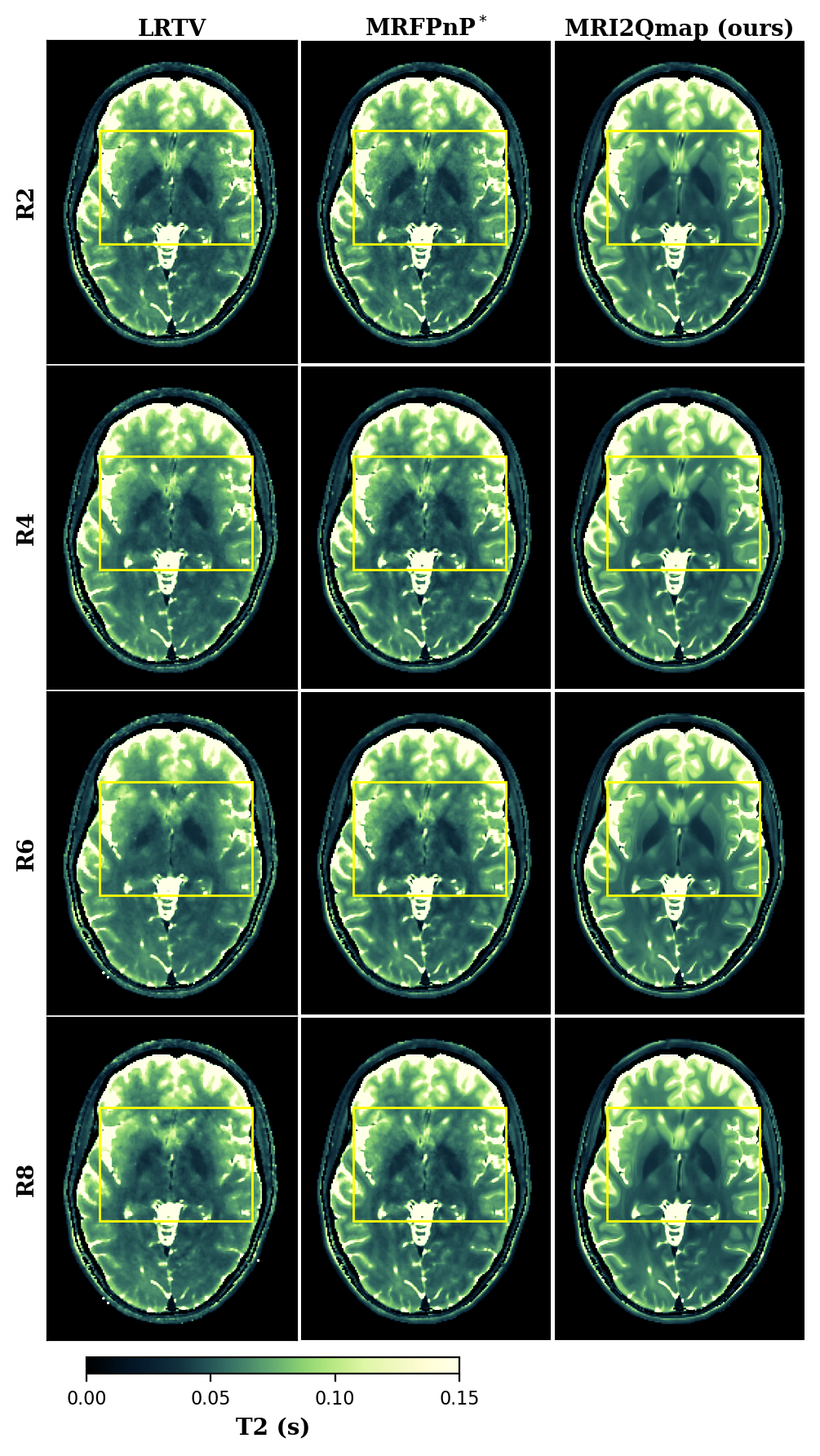}
    \caption{Reconstructed T1 and T2 maps at acquisition acceleration factors $R=2,4,6,8$ (rows) for a representative \invivo\ test image.  
    Relative to baselines ($^*$indicates MRF-trained), \mriq\ exhibits reduced artifacts and improved delineation of deep brain structures (electronic zoom in boxed regions is recommended).}
\label{fig:R2R8_vivo}
\end{figure*}

\end{document}